\documentclass[fleqn,usenatbib]{mnras}
\usepackage{newtxtext,newtxmath}
\usepackage{ mathrsfs,amsmath }

\usepackage[T1]{fontenc}
\usepackage{ae,aecompl}

\usepackage{graphicx}	% Including figure files
\usepackage{amsmath}	% Advanced maths commands
\newcommand{\mue}{$\langle\mu_*\rangle_e$}
\newcommand{\mustar}{$\mu_*$}
\newcommand{\agewr}{$\mathit{Age_{wr}}$}
\newcommand{\zstarwr}{$\mathit{Z_{*wr}}$}
\newcommand{\msunpcsq}{$\mathrm{M}_\odot\,\mathrm{pc}^{-2}$}
\defcitealias{Zibetti:2017aa}{Z17}
\title[``Glocal'' stellar population drivers]{Stellar mass as the ``glocal'' driver of galaxies' stellar population properties}

\author[S. Zibetti \& A. R. Gallazzi]{Stefano Zibetti$^{1}$\thanks{E-mail: stefano.zibetti@inaf.it}
\&
Anna R. Gallazzi$^{1}$\\
$^{1}$INAF-Osservatorio Astrofisico di Arcetri, Largo Enrico Fermi 5, I-50125 Firenze, Italy}
\date{Accepted 2022 February 07. Received 2022 January 28; in original form 2021 December 01}

\pubyear{2022}

\begin{document}
\label{firstpage}
\pagerange{\pageref{firstpage}--\pageref{lastpage}}
\maketitle

\begin{abstract}
The properties of the stellar populations in a galaxy are known to correlate with the amount and the distribution of stellar mass. We take advantage of the maps of light-weighted mean stellar age \agewr~and metallicity \zstarwr~for a sample of 362 galaxies from the integral-field spectroscopic survey CALIFA (summing up to $>600\,000$ individual regions of $\sim 1$~kpc linear size), produced in our previous works, to investigate how these \emph{local} properties react to the \emph{local} stellar-mass surface density \mustar~and to the \emph{global} total stellar mass $M_*$ and mean stellar-mass surface density \mue.
We establish the existence of \emph{i)} a dual \mustar--\agewr~relation, resulting in a young sequence and an old ridge, and \emph{ii)} a \mustar--\zstarwr~relation, overall independent of the age of the regions. The global mass parameters ($M_*$ and, possibly secondarily, \mue) determine the distribution of \mustar~in a galaxy and set the maximum attainable \mustar, which increases with $M_*$. $M_*$ affects the shape and normalization of the local relations up to a threshold mass of $\sim 10^{10.3}\mathrm{M}_\odot$, above which they remain unchanged. We conclude that \emph{stellar mass is a ``glocal'' (i.e. simultaneously global and local) driver of the stellar population properties}. We consider how the local and global mass--age and mass--metallicity relations are connected, and in particular discuss how it is possible, from a single local relation, to produce two different global mass--metallicity relations for quiescent and star-forming galaxies respectively, as reported in the literature. Structural differences in these two classes of galaxies are key to explain the duality in global scaling relations and appear as essential in modelling the baryonic cycle of galaxies.
\end{abstract}


%%%%%%%%%%%%%%%%%%%%%%%%%%%%%%%%%%%%%%%%%%%%%%%%%%

%%%%%%%%%%%%%%%%% BODY OF PAPER %%%%%%%%%%%%%%%%%%

\section{Introduction}
%This is a simple template for authors to write new MNRAS papers.
%See \texttt{mnras\_sample.tex} for a more complex example, and \texttt{mnras\_guide.tex}
%for a full user guide.
%
%All papers should start with an Introduction section, which sets the work
%in context, cites relevant earlier studies in the field by \citet{Others2013},
%and describes the problem the authors aim to solve \citep[e.g.][]{Author2012}.

The stellar mass content of a galaxy has been long recognized as a fundamental \emph{extensive} parameter, that is strongly connected with a number of \emph{intensive} properties, not only of stellar populations (such as age, metallicity, e.g. \citealt{gallazzi+05}, elemental abundance ratios, e.g. \citealt{thomas+05,Gallazzi:2021aa}), but also concerning star formation \citep[e.g.][]{brinchmann+04,Renzini:2015aa}, gas in its multiphase components \citep[e.g.][]{mannucci:2010,saintonge:2017}, dynamics \citep[e.g.][]{falcon-barroso:2019} and morphology \citep[e.g.][]{nair_abraham_2010}. Whether these tight correlations imply a direct physical causation by the stellar mass (e.g. by dynamical effects) or derive from the fact that all these quantities are the results of a galaxy's evolution process is one of the most fundamental yet most complex questions of galaxy astrophysics.

On the way to try and address this problem, one of the long-standing open questions is whether it is stellar mass \emph{density} rather than the \emph{total} stellar mass the main parameter that should enter in these so-called scaling relations. Several authors in the past two decades have investigated this issue
by confronting the impact on galaxy properties of \emph{global} average stellar mass density and of total stellar mass. Since these two quantities are significantly correlated, past investigation has often reached contrasting conclusions (e.g. \citealt{kauffmann+03b} vs. \citealt{bell_dejong00}). In any case, the key role of stellar mass density across a large range of cosmic time is now quite obvious, as already pointed out by \cite{williams:2010}.

With the advent of integral field spectroscopic (IFS) surveys it has become possible to investigate the relations between stellar mass density and the other intensive properties on \emph{local} (i.e. on $\lesssim$kpc) scales, rather than globally. Detailed studies on nearby samples suggest that it is actually the local gravitational potential what makes the stellar mass density such a key parameter in determining the physical properties of galaxies \citep{Scott:2009,Barone:2018,Barone:2020}.
\cite{Gonzalez-Delgado:2014aa}, in particular, analysed the IFS data of 107 CALIFA galaxies and argued that both local and global scales shape the relations between stellar mass (density) and star formation history, with distinct trends applying to discs and bulges. Qualitatively similar conclusions have been recently reached by \cite{Neumann.2021} based on their analysis of MaNGA galaxies.

In this paper we present a systematic investigation of the impact of stellar mass and stellar mass density, considered both on \emph{local} and \emph{global} scales, \emph{on the local stellar population properties}, namely mean light-weighted age and metallicity, based on a sample of 362 CALIFA galaxies \citep{Sanchez:2012aa} observed in IFS. The linear physical scale of ``local'' estimates is dictated by the observational property of CALIFA, i.e. approximately 1 kpc. In particular we consider three different parameters describing the content and distribution of the stellar mass in a galaxy: the \emph{local} (region by region) stellar mass surface density $\mu_*$, the \emph{total} stellar mass $M_*$, and the \emph{global} average stellar mass surface density within one effective radius \mue. First we show how each of these three parameters correlate individually with local mean light-weighted age and metallicity. Then we analyse how these stellar population properties depend on the bivariate combinations of local $\mu_*$ with total $M_*$ and global \mue, respectively. We also discuss how the properties of spatially resolved regions can give rise to global scaling relations and what is the extra information that this approach can bring to the interpretation of such relations in terms of ``baryonic cycle'' \citep[e.g.][]{Peng-Maiolino:2014ab}. In particular we will investigate the origin, in terms of local relations and galaxy structure, of the observed duality in the global mass--metallicity relation, whereby quiescent galaxies display a flatter and higher-normalization relation with respect to the steeper relation followed by star-forming galaxies \citep[e.g.][]{peng_maiolino_cochrane2015,Gallazzi:2021aa}. 

The paper is organized as follows: Sec.\ref{sec:data stelpop} presents the data and the spatially resolved analysis of the stellar population properties, including the stellar mass and mass surface density; Sec. \ref{sec:glocal_mass_corr} analyses the correlations between total stellar mass, mean stellar-mass surface density and local stellar-mass surface density; Sec. \ref{sec:Z_age_MR} presents the spatially resolved relations between stellar mass (density), on one side, and mean stellar age and metallicity, on the other side; Sec. \ref{sec:glocal_drivers} analyses the joint influence of global and local mass (density) on the age and metallicity of stars; in Sec. \ref{sec:discussion} we discuss the results, with particular attention to the implications that they have on the interpretation of the dual mass-metallicity relation for quiescent and star-forming galaxies; our results are summarized and conclusions are given in Sec. \ref{sec:summary}.

Throughout the paper we adopt the cosmology defined by $H_0=70~\mathrm{km~s^{-1}~Mpc^{-1}}$, $\Omega_\Lambda=0.7$, and a flat Universe, consistently with the sample characterization and $V_\mathrm{max}$ corrections (see below) of \cite{Walcher:2014aa}.

\section{Data and stellar population analysis}\label{sec:data stelpop}
We base the present analysis on the database of spatially resolved stellar population properties introduced in 
\citet[][Z17]{Zibetti:2017aa}. We refer the reader to \citetalias{Zibetti:2017aa} for an extended description and summarize here the
basic information.
\subsection{Sample selection and basic data processing}
The initial observational dataset is composed of 394 galaxies that are part of the so-called CALIFA ``Main Sample'' and have available integral field spectroscopy (IFS) from the third and final data release (DR3) of the CALIFA survey \citep{Sanchez:2012aa,Sanchez:2016aa}, over the full optical range ($3700\,$\AA ~to $7140\,$\AA, so-called ``COMBO'' cubes). The average spectral resolution is $6\,$\AA~FWHM and the effective spatial resolution is $\sim 2.57\arcsec$ FWHM. The IFS is complemented by SDSS DR7 \citep{SDSS,SDSS_DR7} broad-band $ugriz$ imaging.
These 394 galaxies are effectively a random selection of the diameter-selected CALIFA ``mother sample'', as described in \cite{Walcher:2014aa}, and include all galaxies that are part of the CALIFA ``Main Sample'' and are observed in the COMBO setup\footnote{We do not consider the extension samples, although they are part of the CALIFA DR3, since they would compromise the representativeness of the sample.}, except two galaxies that could not be analyzed\footnote{As reported in \citetalias{Zibetti:2017aa}, we excluded UGC\,11694 because of a very bright star near the centre, which contaminates a significant portion of the galaxy’s optical extent, and UGC\,01123 because of artefacts in the noise spectrum that make the stellar population analysis unreliable in a significant portion of spaxels.}. A representativeness $>95\%$ is obtained in the stellar-mass range $9.65<\log(M_*/\mathrm{M}_\odot)<11.44$ for the mother sample in the nominal redshift range $0.005 <z< 0.03$. The so-called ``$V_{\mathrm{max}}$'' volume corrections, which take into account also the large scale density fluctuations ($V^\prime_{\mathrm{max}}=\delta \times V_{\mathrm{max}}$, see Sec. 4.1 of \citealt{Walcher:2014aa}), provide the statistical weight of each galaxy and allow us to cover the range from $\sim 10^9 \mathrm{M}_\odot$ to $\sim10^{12}\mathrm{M}_\odot$. 
We further filter out galaxies that are likely to be either ongoing or very recent mergers, based on the morphological classification performed by the CALIFA collaboration \citep[see sec. 6.4 of][]{Walcher:2014aa}, in order to focus on (quasi) equilibrium systems. This leaves us with a final sample of 362 galaxies.

CALIFA datacubes and SDSS images are pre-processed to
achieve an accurate spatial match in terms of both registration and effective resolution. In each spaxel we want to extract a set of 
spectroscopic indexes and broad-band photometric fluxes that we use as observational constraints to estimate the relevant stellar population properties.
A crucial step in the analysis is the decoupling of the stellar continuum from the nebular emission lines, which is not only essential
to correct the Balmer absorption indices for the infill by the recombination lines, but is also required to clean photometric fluxes
from the contamination of lines that can reach very high equivalent width in individual spaxels. To this goal, we adopt the following procedure. First we use \texttt{pPXF} \citep{Cappellari_pPXF} to determine the stellar kinematics on the spectra where regions 
potentially affected by emission lines are masked. Then \texttt{GANDALF} \citep{Sarzi:2006aa} is run with the stellar kinematics obtained
by \texttt{pPXF} to perform the actual decoupling. Index measurements are finally performed on the pure stellar spectra (i.e. original spectra minus best fit emission lines) at each spaxel and
corrections for the emission lines are applied to the photometric fluxes.

We note that a reasonably high signal-to-noise ratio (SNR) is required both for the stellar-nebular decoupling and for obtaining significant index measurements. Therefore, in first place, we apply a surface brightness cut and consider only spaxels brighter than $22.5~\mathrm{mag~arcsec}^{-2}$ in $r$-band. Then, before running the kinematic analysis and the emission line decoupling, the IFS cubes and the corresponding broad-band images are processed with the algorithm of spatial adaptive smoothing \texttt{adaptsmooth}. This was originally designed for images by \cite{ZCR09} and \cite{adaptsmooth} and subsequently extended to IFS in order to enhance the signal-to-noise ratio in the low surface-brightness spaxels \citepalias[see][for more details]{Zibetti:2017aa}. Essentially, we replace the spectra of low-SNR spaxels with the median-average of a sufficiently large number of surrounding spaxels, until a SNR of 15 per \AA~(20 per spectral pixel) on the averaged spectrum is reached, up to a maximum distance of 5\arcsec. While it is worth noting that the adaptive smoothing inevitably introduces spurious spatial correlations between spaxels, we argue that this effect is minimized with respect to the popular Voronoi binning approach, as we discuss in Appendix \ref{app:azmooth3}. In there, we also show that the main results of this work persist even if the analysis is restricted to the spaxels that do not require any smoothing.

For the stellar population analysis, we retain only spaxels that provide a minimum SNR of 7.5 per \AA~(10 per spectral pixel) in the smoothed cube. After further excluding all spaxels that are contaminated by foreground sources or artifacts, or for which any of the procedures described in the previous paragraph fail, our final sample is made of $601\,282$ spaxels, corresponding to $90.8\%$ of the total spaxels above the surface brightness cut.\\

\subsection{Stellar population parameters}\label{sub:SPpars}
The mean $r$-band-light-weighted stellar age and metallicity, along with the local stellar mass surface density, are estimated with the Bayesian method introduced by \cite{gallazzi+05}.
A vast library of $500\,000$ synthetic spectra are produced according to randomly generated star-formation and chemical-enrichment histories
and two-component dust attenuations \citep[\'a la][]{charlot_fall}. Specifically, SFHs are modelled as a continuum component $\mathit{SFR}(t)\propto\frac{t}{\tau}e^{-\frac{t^2}{2\tau^2}}$ \citep{Sandage:1986aa}, thus allowing for both rising and declining (phases of the) SFH, with superimposed bursts in random numbers (including zero) and with random age and intensity. For the chemical enrichment we consider only solar-scaled elemental abundance ratios and let the total metallicity evolve as a function of formed stellar mass according to a pseudo-``leaking box'' model \citepalias[eq. 3 of][]{Zibetti:2017aa}. All models are based on the simple stellar populations (SSPs) produced by \cite[BC03]{BC03} stellar population synthesis code in its 2016 version \citep{Chevallard:2016aa},  using the MILES spectral libraries \citep{Sanchez-Blazquez:2006aa,Falcon-Barroso:2011aa}, and assuming a universal \cite{chabrier03} initial mass function (IMF).
The mean $r$-band-light-weighted stellar age \agewr~and metallicity \zstarwr~of each model are defined on the linear quantities as follows:
\begin{equation}\label{eq:agewr}
\begin{split}
\mathit{Age_{wr}} & =\frac{\int\limits_{t=0}^{t_0} dt~(t_0-t)~\mathrm{SFR}(t)~\mathscr{L}_r^\prime(t)}{\int\limits_{t=0}^{t_0} dt~\mathrm{SFR}(t)~\mathscr{L}_r^\prime(t)}= \\
& \hspace{3cm} =\frac{\int\limits_{t=0}^{t_0} dt~(t_0-t)~\mathrm{SFR}(t)~\mathscr{L}_r^\prime(t)}{L_r}
\end{split}
\end{equation}
\begin{equation}\label{eq:Zwr}
\mathit{Z_{*wr}}=
\frac{\int\limits_{t=0}^{t_0} dt~Z_*(t)~\mathrm{SFR}(t)~\mathscr{L}_r^\prime(t)}{L_r} 
\end{equation}
Here $t$ is the time since the on-set of the SFH, $t_0$ is the epoch of observation and $\mathrm{SFR}(t)$ is the star formation rate as a function of time. $\mathscr{L}_r^\prime(t)$ is the $r$-band luminosity per unit \emph{formed} stellar mass output by the stars of age $(t_0-t)$ and decreased by the effective dust attenuation. $L_r\equiv \int\limits_{t=0}^{t_0} dt~\mathrm{SFR}(t)~\mathscr{L}_r^\prime(t)$ is total $r$-band luminosity at $t_0$.

The estimates of these two physical quantities and of the stellar mass surface density \mustar~are based on the comparison between models and observed data for five spectral absorption indices and five photometric bands. The five indices are the same as the set originally defined in \cite{gallazzi+05}: three (mostly) age-sensitive indices ($\mathrm{D4000_n}$, $\mathrm{H\beta}$, and $\mathrm{H\delta_A}+\mathrm{H\gamma_A}$) and two (mostly) metal-sensitive composite indices that show minimal dependence on $\alpha$-element abundance relative to iron-peak elements ($[\mathrm{Mg_2Fe}]$ and $[\mathrm{MgFe}]^\prime$)\footnote{ Individual Fe or Mg indices cannot be used for a comparison with models that do not include variable abundance ratios, like in our case, where we adopt solar-scaled models, i.e. elemental abundance ratios that are fixed to the solar neighbourhood values, rescaled by total metallicity. For all galaxies/regions for which the elemental abundance ratio is different from solar, these individual element indices \emph{i)} would be in tension with each other, chiefly the Fe-peak vs $\alpha$-element indices, but also with other indices (such as the Balmer) that are secondarily dependent on metallicity, and \emph{ii)} would lead to a biased estimate of the total metallicity (and hence of the age, via the well-known age-metallicity degeneracy). This is the reason why, since \cite{gallazzi+05}, we rely on Mg-Fe composite indices that display minimal dependence on the $\alpha$/Fe element ratio, as demonstrated by \cite{thomas+03} and \cite{BC03}.}. The photometric fluxes in the five SDSS bands ($ugriz$) provide the normalization (hence the stellar mass) and the sensitivity to dust attenuation, and complement the constraints from the indices.
The likelihood for the data given each $i$-th model is computed via standard chi-squared on the observable quantities, as $\mathcal{L}_i\propto\exp(-\chi_i^2/2)$, after taking out the normalization factor (i.e. the best-fit stellar mass for the $i$-th model) via chi-squared minimization.
By marginalizing the probability distribution for the entire library over all other parameters, we obtain the posterior probability distribution (pPDF) for \agewr, \zstarwr~and \mustar, respectively. The value corresponding to the median of the pPDF is taken as fiducial estimate. As uncertainty we take half of the range between the 16th and 84th percentiles (i.e. $\pm1\sigma$  for a gaussian distribution).

\begin{figure}
	\includegraphics[width=0.5\textwidth]{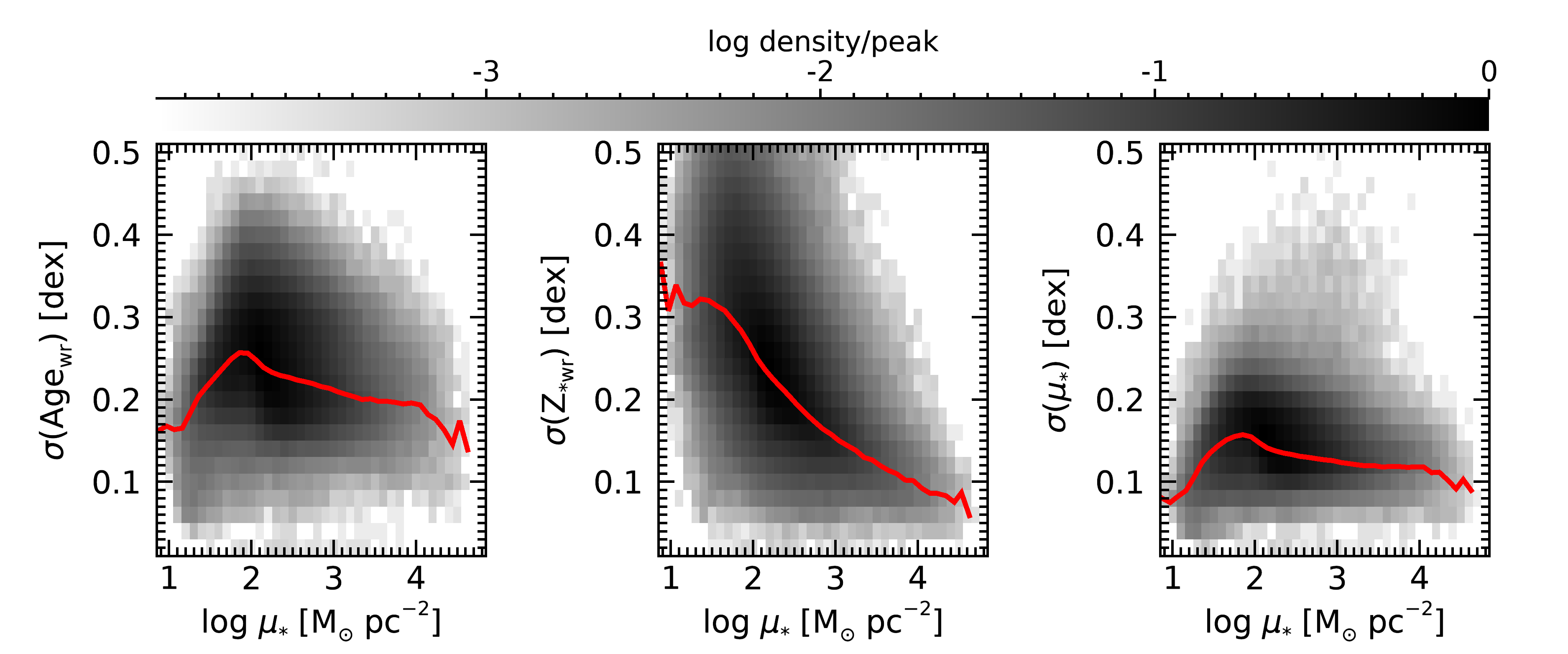}
	
    \caption{Distributions of errors on the spatially-resolved stellar population properties. The color scale displays the density of spaxels on the plane, normalized to the peak. The red line displays the running median.}
    \label{fig:errors}
\end{figure}

Typical uncertainties on both \agewr~and \zstarwr~are around $0.2~\text{dex}$, although with systematic variations as a function of \mustar. Specifically, the median error on \agewr~has a peak of $\sim 0.25~\text{dex}$ around $\log \mu_*/\mathrm{M}_\odot \mathrm{pc}^{-2} \sim 2$ and decreases below $\sim 0.2~\text{dex}$~at larger and smaller \mustar. The error on \zstarwr~monotonically decreases from $\sim 0.3~\text{dex}$~at $\log \mu_* \sim 1$ to $\lesssim 0.1~\text{dex}$ at $\log \mu_* \sim 4$. The median error on \mustar~ranges between $0.1~\text{dex}$ (at the highest \mustar) and $0.15~\text{dex}$. The distributions of errors on the spatially-resolved stellar population properties as a function of \mustar~are shown in Fig. \ref{fig:errors}, with the red lines displaying the running median relations.

For each galaxy we consider the ``total'' stellar mass $M_*$ taken from \cite{Walcher:2014aa} based on optical photometry alone, which provides typical uncertainties of $0.1$--$0.15~\text{dex}$.
The average surface stellar mass density within one effective radius, \mue, is computed by integrating the stellar mass surface density (median-likelihood fiducial value in each spaxel) within the elliptical isophotal aperture that encloses $50\%$ of the total $r$-band flux \citep[according to the growth curve analysis of][]{Walcher:2014aa}, which is then divided by the area of the ellipse.\\

In the parameter distributions that we present in the following sections, each spaxel is weighted by the product of its physical projected area on plane of the sky (in $\mathrm{pc}^2$ as derived from its parent galaxy's redshift and pixel scale) and of the statistical weight $1/V^\prime_{\mathrm{max}}$ (introduced at the beginning of this section).
This weighting scheme allows us to reduce the bias against the most massive galaxies, on one hand, and the extended, low-surface-brightness galaxies on the other hand. However, the sample of regions is affected by systematic selection effects which result from the surface-brightness and SNR cuts that we must apply in order to obtain meaningful measurements of the stellar population properties. In fact, a fixed surface brightness limit translates into a variable surface mass-density limit as a function of age mainly, and of metallicity and dust attenuation secondarily. The estimated $M/L$ ratio differs by almost one order of magnitude between the oldest and the youngest stellar populations probed in this study. This results in a completeness cut at a few $10\,\mathrm{M}_\odot\,\mathrm{pc}^{-2}$ for the youngest regions, while for the oldest ones the completeness limit is at almost $100\,\mathrm{M}_\odot\,\mathrm{pc}^{-2}$.

\section{Correlations between global and local stellar mass (density)}\label{sec:glocal_mass_corr}
\begin{figure*}
	\includegraphics[width=\textwidth]{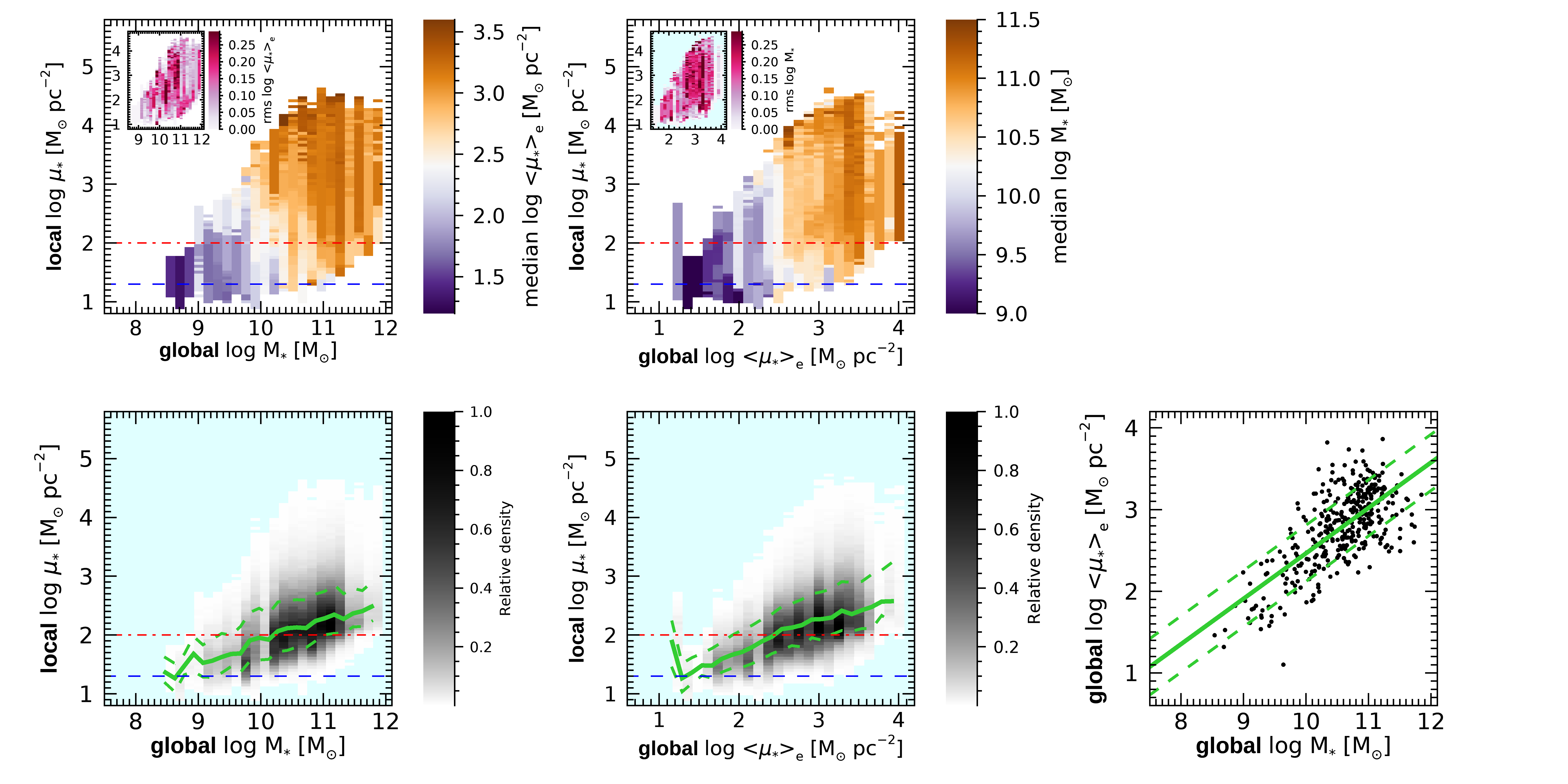}
    \caption{Correlations among the three different stellar mass (density) parameters. The panels in the bottom row display the bivariate distributions. The bottom left and mid panels show the distributions for the individual regions, weighted by projected area and corrected for the survey available volume (grey shading), with running median (solid line) and 16th and 84th percentiles (dashed lines) plotted in green. The rightmost panel displays
    the correlation between the two global quantities, $M_*$ and \mue, with each point representing one galaxy. The robust linear fit to the points is plotted with the solid green line, with the two dashed lines representing $\pm 1\,\sigma$ about the fit. The top panels illustrate the median values (and the scatter, in the insets) of \mue at fixed $(M_*,\mu_*)$ and of $M_*$ at fixed $\left(\langle\mu_*\rangle_e,\mu_*\right)$, respectively.
    The red dot-dashed line and the blue dashed line indicate the approximate completeness limit for the old-ridge and the young-sequence regions, respectively (see Sec. \ref{sub:SPpars}).}
    \label{fig:glocal_mass_corr}
\end{figure*}

Before analyzing how the different global and local stellar mass (density) impact on the local age and metallicity, it is worth considering how these masses/mass densities are mutually correlated. The bottom panels of Fig. \ref{fig:glocal_mass_corr} display the three projections of the \mustar--\mue--$M_*$ space. The strongest correlation is found between the global quantities $M_*$ and \mue, which scale as a powerlaw of slope $0.55$, yet with a significant scatter of $0.34\,\mathrm{dex}$ in \mue~at fixed $M_*$. This is a well known correlation \citep[e.g.][]{blanton_etal03}, whereby more massive/luminous galaxies have higher mean surface mass-density/brightness. 

The correlations between \mustar~and $M_*$ and between \mustar~and \mue, respectively, are much milder, although highly significant: the powerlaw slope is $\approx 0.2$
and the typical scatter is $\approx 0.4$--$0.5\,\mathrm{dex}$. While the correlation between \mustar~and \mue~is expected by construction, the correlations between $M_*$ and the surface mass-density (either local or globally averaged) reflect the non-trivial link between total stellar mass and galaxy structure. Part of the \mustar--$M_*$ relation, however, is due to the surface brightness cut applied to our data and the correlation between $M_*$ and stellar age and metallicity. In fact, in more massive galaxies the given surface brightness cut corresponds to a higher \mustar~cut because of the higher stellar $M/L$ ratio of the older and more metal-rich stellar populations.

One important property emerging from these plots is that regions belonging to galaxies of $M_*\lesssim 10^{10}\,\mathrm{M_\odot}$ (or \mue$\lesssim10^2 \mathrm{M_\odot\,pc^{-2}}$) never reach \mustar~above $10^3 \mathrm{M_\odot\,pc^{-2}}$. In other words, \emph{the maximum surface mass-density that a galaxy can reach is a strong function of its total stellar mass}.

The top panels of Fig. \ref{fig:glocal_mass_corr} display the trends of \mue~as a function of $M_*$ and \mustar, and of $M_*$ as a function of \mue~and \mustar\, (left and right panel, respectively). The main panels represent the median value at a given $(x,y)$, while the insets display the corresponding r.m.s. The three quantities are positively correlated one another, yet the r.m.s. exceeds $0.15\,\mathrm{dex}$ in most positions of the planes.

These plots also illustrate that, despite the intrinsic correlations, our data cover a large enough dynamical range to study the correlations of \agewr~and \zstarwr~with each of the local and global mass (density) parameters independently, while controlling for the others.

\section{Spatially resolved mass--age and mass--metallicity relations}\label{sec:Z_age_MR}
In this section we study the bivariate relations between stellar mass and stellar mass density as independent variables, on one side, and mean light-weighted stellar age \agewr~and metallicity \zstarwr~as dependent variables, on the other side. 

\subsection{Overall mass--age relations}\label{sub:mass_age_all}
\begin{figure*}
	\includegraphics[width=\textwidth]{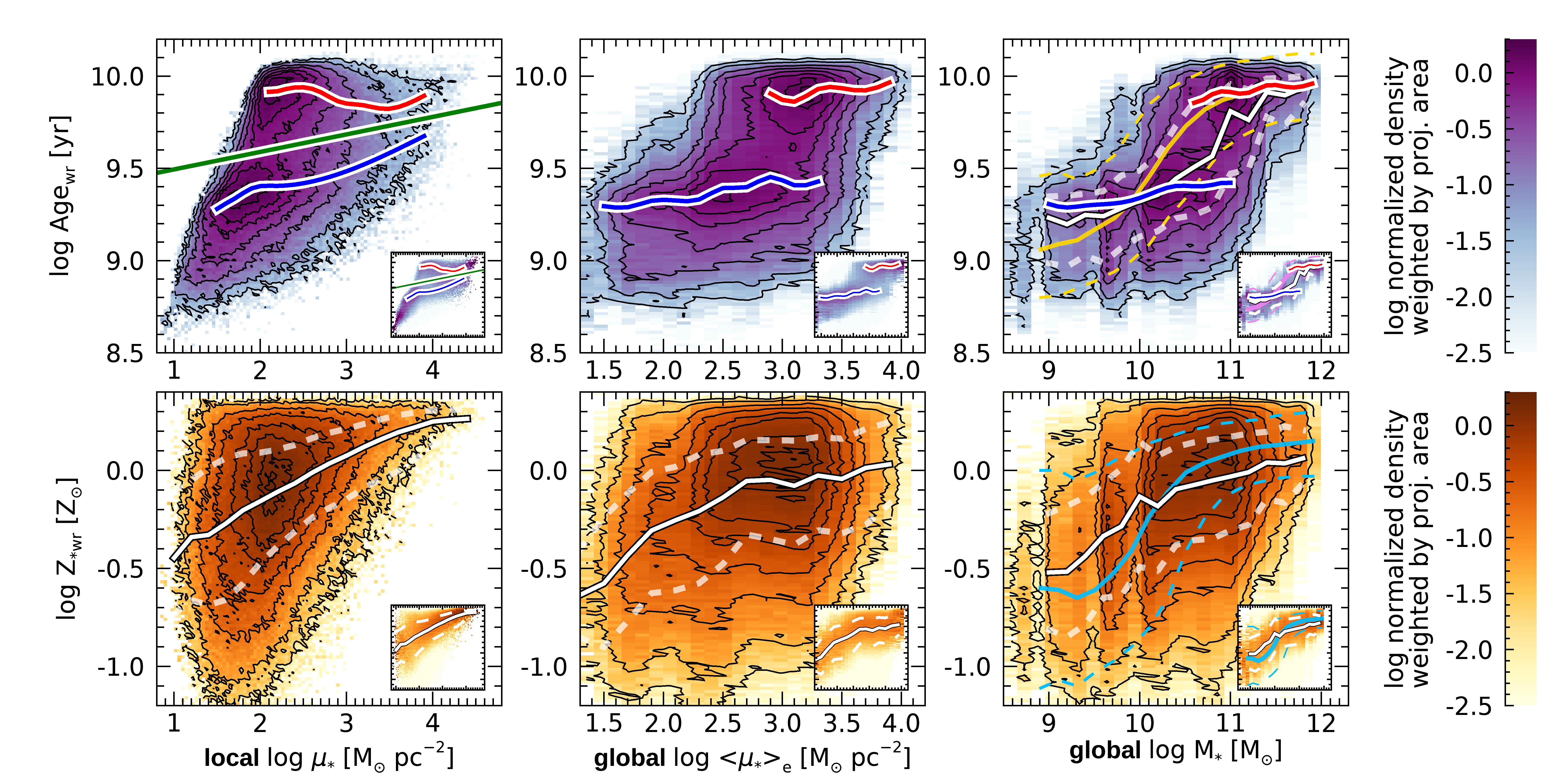}
	
    \caption{Distributions of regions weighted by physical projected area and normalized to the peak, in the \agewr--mass (density) planes (\emph{top row}) and in the \zstarwr--mass (density) planes (\emph{bottom} row). In each row, from left to the right, the $x$ axis represents the \emph{local} stellar mass density \mustar, the \emph{global} mean stellar mass density within 1 effective radius \mue, and the \emph{global} stellar mass $M_*$. Logarithmically-spaced isodensity contours are over-plotted. The insets of each panel represent the corresponding distributions after normalizing the integral of the distribution in each $x$-bin.
    In the age--mass panels the blue and the red lines trace the positions of the peaks of the distribution at fixed mass (density). The green line in the \mustar--\agewr~panel displays the divide between ``old-ridge'' and ``blue-sequence'' regions. In the metallicity--mass panels and in the \agewr--$M_*$ panel, the white lines display the median (solid) and the 16th and 84th percentiles (dashed). The yellow lines in the \agewr--$M_*$ and the light blue lines in the \zstarwr--$M_*$ panels are the global relations for SDSS galaxies from \citet{gallazzi+05}.}
    \label{fig:ageZ_mass_relation}
\end{figure*}

In the top row of Fig. \ref{fig:ageZ_mass_relation}, the three panels display the distribution of regions weighted by physical projected area in the 2D parameter space defined by the local \agewr~(on the $y$ axis) and the three different quantities related to stellar mass (on the $x$ axis), respectively: \textit{i)} the \textit{local} stellar-mass surface density of the specific region, \mustar~(left panel), \textit{ii)} the \textit{global} average stellar-mass surface density within $R_e$, \mue~(mid panel), and \textit{iii)} the \textit{global} total stellar mass of the galaxy, $M_*$ (right panel). In all three cases a bimodality in stellar age is visible. 
This is made more evident by plotting the distributions obtained after re-normalizing in each $x$-axis bin, as shown in the panel insets, so to avoid the visual bias due to the non-uniform distribution of regions in mass (density). To further highlight this bimodality, we locate the two \agewr~peaks of the distributions in broad mass (density) bins and connect them with the red and blue lines, respectively.
It is particularly intriguing to see that the age bimodality is present at any value of \mustar, and determines, as a function of \mustar, an ``old ridge'' of regions with almost constant \agewr$\sim 10^{9.8}$ Gyr (red lines), and a ``young sequence'' along which we find regions of younger age, decreasing as \mustar~decreases (blue lines). This result was already noted in \citetalias{Zibetti:2017aa}. The width of the peaks along the \agewr~axis is comparable with the typical errors on \agewr~(see Fig. \ref{fig:errors}), thus implying that the intrinsic dispersion of the two sequences is typically smaller than $\approx 0.2\, \mathrm{dex}$.
We define the division between old ridge regions and young sequence regions in the \mustar--\agewr~plane by fitting a straight line through the ``green valley'' of \agewr~minima as a function of \mustar:\footnote{More specifically, we consider the distributions in \agewr~after binning all regions in bins of 0.25 dex in \mustar. In each distribution we find the minimum with a simple inverse-peak finder algorithm. Finally we perform a simple linear least-square fit between the central \mustar~of the bins and the corresponding log(\agewr) of the minimum. This linear fit defines the ``green valley'' division line.}
\begin{equation}\label{eq:green_valley}
    \log(\mathit{Age_{wr,GV}}/[\mathrm{yr}])=0.095\,\log(\mu_*/[\mathrm{M}_\odot\,\mathrm{pc}^{-2}]) + 9.40
\end{equation}
This is marked with a straight green line in the top left panel of Fig. \ref{fig:ageZ_mass_relation}.

Interestingly, using the global parameters \mue~and $M_*$ instead of the local \mustar, young and old regions appear to be more segregated: the old ridge extends essentially only over the high $M_* \gtrsim 10^{10.3}\mathrm{M}_\odot$ range and over the high \mue$\gtrsim 10^{2.8}$\msunpcsq~range; young sequence regions, on the contrary, \textit{globally} avoid the most massive and dense galaxies. Moreover, the young sequence is considerably flatter as a function of \mue~or $M_*$ than as a function of \mustar. This is a possible indication that the fundamental correlation is with the local \mustar, while correlations with the \emph{global} mass and mass density are of second order, as if regions are reshuffled along the $x$-axis when going from \mustar~to \mue~or $M_*$.

The \agewr~vs $M_*$ distribution can be directly compared with the analogous distribution for SDSS galaxies, where the \textit{local} \agewr~is replaced by the \textit{global} integrated mean light-weighted age of \citet[][Fig. 8]{gallazzi+05}. The median relation and the 16th and 84th percentiles from \citet{gallazzi+05} are overplotted as yellow lines (solid for the median and dashed for the percentiles). Similarly, we have derived the median and the percentiles in each $M_*$ bin of our distribution and overplotted them in white. The spatially-resolved and the integrated relations share the same qualitative trend, whereby stellar age flattens out at low ages going to low $M_*$, and at high ages going to high $M_*$, respectively. The transition from one regime to the other occurs between $10^{10}$ and $10^{11}\,\mathrm{M}_\odot$. However, there are relevant quantitative differences. First of all, in the integrated relation the transition is much sharper and occurs around the well-known critical mass of $10^{10.3}\,\mathrm{M}_\odot$, approximately where the old ridge fades. Moreover, in the low-mass end, the median integrated relation reaches some $0.2\,\text{dex}$ below the spatially resolved one. This may suggest that in low-mass galaxies, regions characterized by young and metal poor stars\footnote{The presence of young regions with low metallicity is expected if recent star formation has been caused by ``fresh'' gas infalling from outside the galaxy, causing effectively a metal dilution. This has been observed in real galaxies, e.g. \cite{Cresci:2010aa}.}, which represent only a fraction of a galaxy's mass and surface, outshine the rest of the galaxy \citep[see e.g.][]{ZCR09,Sorba.2018}, thus biasing low the integrated estimate of the age and of the metallicity as well. The median relation for the individual regions is also significantly below the SDSS integrated relation between $10^{10.5}$ and  $10^{11}\,\mathrm{M}_\odot$. This may possibly reflect a bias induced by the limited fibre aperture in the SDSS for early type spirals, which dominate this mass range \citep[e.g.][]{nair_abraham_2010}. The light of the bulge of these galaxies is most likely prevalent in their SDSS spectra, thus leading to an overestimate of the overall \agewr~with respect to the median of the regions over the entire galaxy.

\subsection{Overall mass--metallicity relations}\label{sub:mass_Z_all}
We repeat the analysis presented in the previous section now considering the mean light-weighted metallicity \zstarwr~instead of \agewr. The results are shown in the bottom row of plots in Fig. \ref{fig:ageZ_mass_relation}. The \zstarwr~vs. stellar mass (density) distributions do not display any sign of bimodality, contrary to what we found for \agewr.

On average, we find a monotonic increase of \zstarwr~with local \mustar~(bottom left panel). Despite the visual impression of a very scattered relation from the bivariate distribution, by looking at the percentiles and at the distribution after normalizing in each \mustar~bin (see inset), a well defined relation emerges. The scatter around the median relation is roughly consistent with the typical errorbars on \zstarwr, going from $\sim0.3\,\mathrm{dex}$ in the low-\mustar~end to $\sim0.1\,\mathrm{dex}$ at the highest densities. By simple considerations about combining error and intrinsic scatter in quadrature, we can conclude that the intrinsic scatter around the median \mustar--\zstarwr~relation is of the order of $0.1\,\mathrm{dex}$ \emph{at most}. A similar relation was shown by \cite{Zhuang:2019}, using the same estimate of \zstarwr~as in this work, but independent estimates of \mustar~from on an orbit-based dynamical analysis.

The relations of \zstarwr~with the \emph{global} mass (density) are overall flatter and more scattered, especially in the intermediate/high mass (density) regime. As in the case of \agewr, this suggests that the primary correlation is between the local \mustar~and local \zstarwr, while correlations with \emph{global} mass and mass density are secondary.

We further compare the median relation between local \zstarwr~and galaxy $M_*$ with the analogous global mass-metallicity relation of \cite{gallazzi+05}, in the bottom right panel of Fig. \ref{fig:ageZ_mass_relation}: white lines denote the local relation, while light blue lines are the global relation, solid for the median and dashed for the 16th and 84th percentiles. Similarly to what we noted for \agewr, also in this case, by comparison to the local relation, the global relation displays a much sharper transition between the low-mass (density) -- low-metallicity regime and the high-mass (density) -- high-metallicity regime. The local relation smoothly decreases its slope from low mass to high mass in a roughly monotonic way, while the global relation is characterized by an S-shape.\\
At masses above few $10^{10}\,\mathrm{M}_\odot$ the median \zstarwr~of the individual regions is about $0.1\,\mathrm{dex}$ below the median global relation. The opposite applies at lower masses. These discrepancies can be explained qualitatively if we consider the steep metallicity gradients and the \mustar--\zstarwr~relation that are observed in massive (mainly early-type) galaxies \citep[e.g.][]{Zibetti:2020aa}. On the one hand, we can expect that regions of high \mustar~(hence high surface brightness) dominate the light in integrated spectra and result in high values of light-weighted metallicity. As opposed, lower metallicity regions at lower \mustar~dominate by number (rather by light) and bias low the median local relation. On the other hand, SDSS spectra are affected by the limited fibre aperture and might miss part of the light of the low-metallicity external regions. Furthermore, in the integrated spectra of low-mass galaxies, we might be seeing the effect of young metal-poor regions outshining the more quiescent (and metal-richer) ones, as noted in the previous paragraph. This might help explain why global light-weighted metallicities are on average lower than the median of the individual regions in the low mass regime. 

\subsection{The mass--metallicity relations for the old ridge and the young sequence regions}\label{sub:MZR_oldyoung}
In this Section we split the sample of regions into two subsamples, corresponding to the young sequence and the old ridge, respectively. They are defined based on the ``green valley'' line given in Eq. \ref{eq:green_valley} and plotted in green in the top left panel of Fig. \ref{fig:ageZ_mass_relation}. As expected, as far as \agewr~is concerned, the two subsamples reproduce the same bimodal distribution seen in the \mustar--\agewr~plane also in the planes of \emph{global} mass (density) vs. age, although with a slightly smoother separation, due to the loose correlation between \mustar, on one side, and \mue~and $M_*$, on the other side.

\begin{figure*}
	\includegraphics[width=\textwidth]{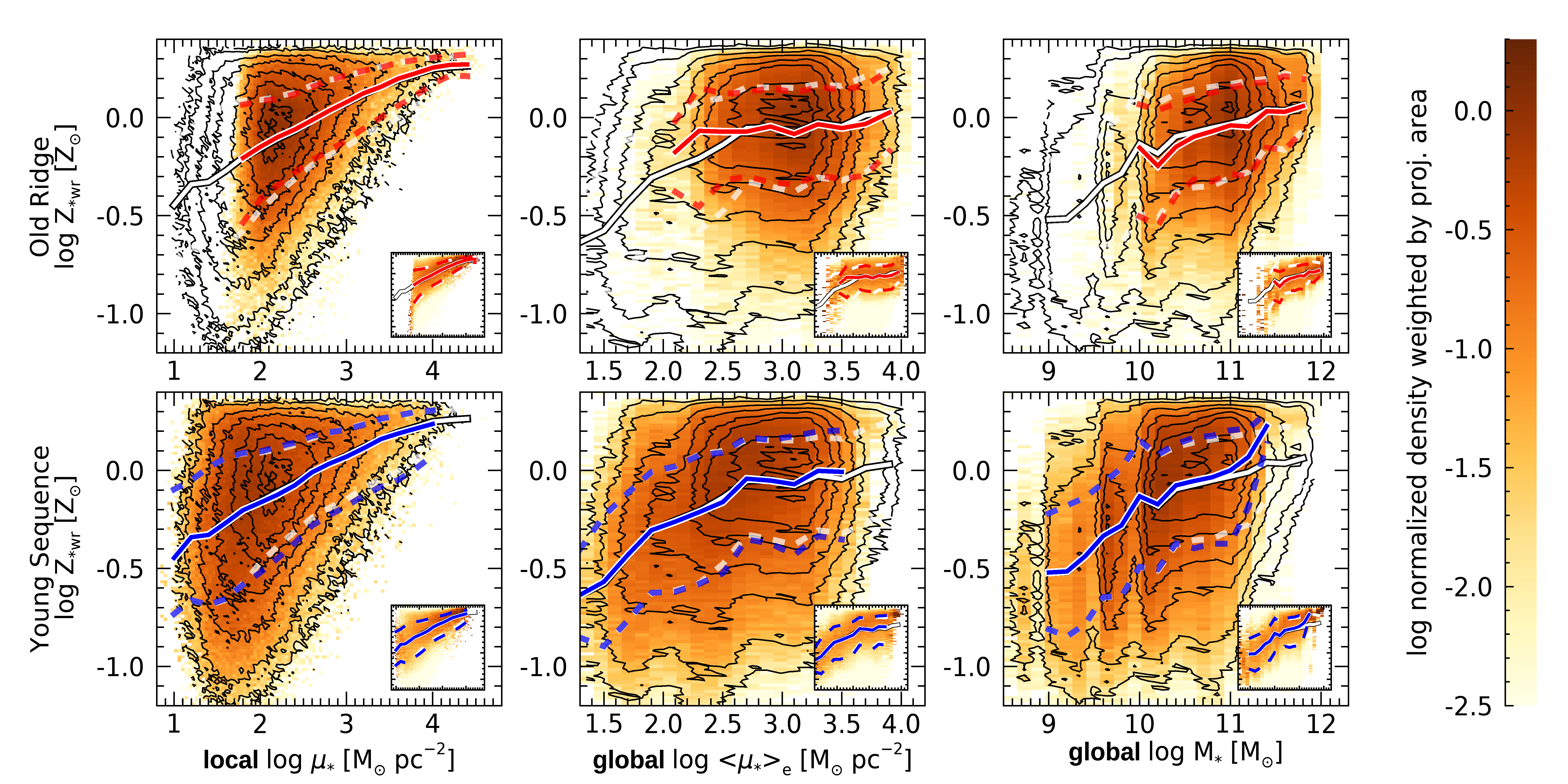}
	
    \caption{Distributions of regions weighted by physical projected area and normalized to the peak, in the \zstarwr--mass (density) planes, for ``old-ridge'' regions (\emph{top} row) and ``young-sequence'' regions (\emph{bottom} row) separately. In each row, from left to the right, the $x$ axis represents the \emph{local} stellar mass density \mustar, the \emph{global} mean stellar mass density within 1 effective radius \mue, and the \emph{global} stellar mass $M_*$. Over-plotted contours represent the logarithmically-spaced isodensity levels for the regions altogether (same as in Fig. \ref{fig:ageZ_mass_relation}). The insets of each panel represent the corresponding distributions after normalizing the integral of the distribution in each $x$-bin.
    The white lines display the median (solid) and the 16th and 84th percentiles (dashed) for all regions (same as in Fig. \ref{fig:ageZ_mass_relation}). The red and blue lines display the corresponding median and percentiles for the `old-ridge'' and ``young-sequence'' regions respectively.}
    \label{fig:Z_mass_relation_bluered}
\end{figure*}

What is most interesting are the distributions of \emph{local} \zstarwr~vs. \emph{local} and \emph{global} mass (density) for young and old regions, respectively. In Fig. \ref{fig:Z_mass_relation_bluered} we show the distributions in the same format as in the bottom panels of Fig. \ref{fig:ageZ_mass_relation}, but using only old-ridge regions in the top row and only young-sequence regions in the bottom row. For reference, we report density contours for the full sample and the median relations and percentiles as white lines. The red and blue lines display the median (solid) and 16th and 84th percentiles (dashed) for the two subsamples, respectively. Quite unexpectedly, the fully \emph{local} \mustar--\zstarwr~relations for the young and the old regions are to large extent indistinguishable from each other and from the overall relation, both in terms of median relation and of scatter. Except for the higher \mustar~limit that applies to old-ridge regions (as a consequence of the fixed SB limit and the higher $M/L$ ratio typical of old stellar populations, see Sec. \ref{sub:SPpars}), the only appreciable difference between the two subsamples is a slightly larger scatter for the young-sequence regions with respect to the old-ridge ones. This may be partly driven by typically larger uncertainties on metallicity for younger stellar populations.

For old-ridge regions, the \mue--\zstarwr~relation is almost flat everywhere for $\log \langle\mu\rangle_e\gtrsim 2.2$, where most of the old-ridge regions live. Young-sequence regions, on the contrary, populate the full range of \mue~and display a clear trend for increasing \zstarwr~at increasing \mue, although the trend flattens out at $\log \langle\mu\rangle_e\gtrsim 2.7$. In the intermediate overlapping range, the two subsamples do share, however, very similar distributions in \zstarwr. A very similar picture emerges considering $M_*$ instead of \mue. It is interesting to note that, at fixed $M_*$, old-ridge regions have a median \zstarwr~consistent with or even a few 0.01 dex \emph{lower} than young-sequence regions\footnote{Although small, a difference of a few 0.01 dex in the median should be considered significant as the running medians are computed adopting bins of 0.2 dex in $M_*$, thus encompassing approximately 30,000 regions per bin per subsample (young-sequence or old-ridge). For a typical random error of 0.2 dex in \zstarwr, the error on the median is expected to be of the order of $0.2~\text{dex} / \sqrt{30,000} \sim~0.001~\text{dex}$. On the other hand, a significant contribution to the fluctuations in the median may arise from the limited galaxy sample. In fact, considering $M_*$ bins of 0.2 dex we have roughly 40 galaxies per bin and therefore the random inclusion or exclusion of a galaxy can affect the region sample at a few percent level and move the median line accordingly. For this reason, the significance of a systematic offset of a few 0.01 dex in \zstarwr~is hard to firmly establish.}.
%\begin{figure*}
%	\includegraphics[width=\textwidth]{figures_prep/age_vs_mass_bluered_6pan.png}
%	
%    \caption{Distribution of regions }
%    \label{fig:age_mass_relation_bluered}
%\end{figure*}

\section{Local and global mass (density) drivers}\label{sec:glocal_drivers}
In the previous sections we have seen that \emph{i)} \emph{local} and \emph{global} mass (densities) are partly correlated and \emph{ii)} age and metallicity are strongly affected by the local stellar mass density \mustar, but also by \emph{global} mass $M_*$ and average mass density \mue. The question we want to address in this Section is whether the dependence of local age and metallicity on global mass (density) is simply inherited from their correlations with local \mustar\, or, on the contrary, the local \mustar--\agewr\, and \mustar--\zstarwr\, relations are affected or modulated by global properties. To this goal we study how \agewr~and \zstarwr~depend on the local \mustar~in different bins of the global parameters, $M_*$ and \mue, respectively.
\begin{figure*}
	\includegraphics[width=\textwidth]{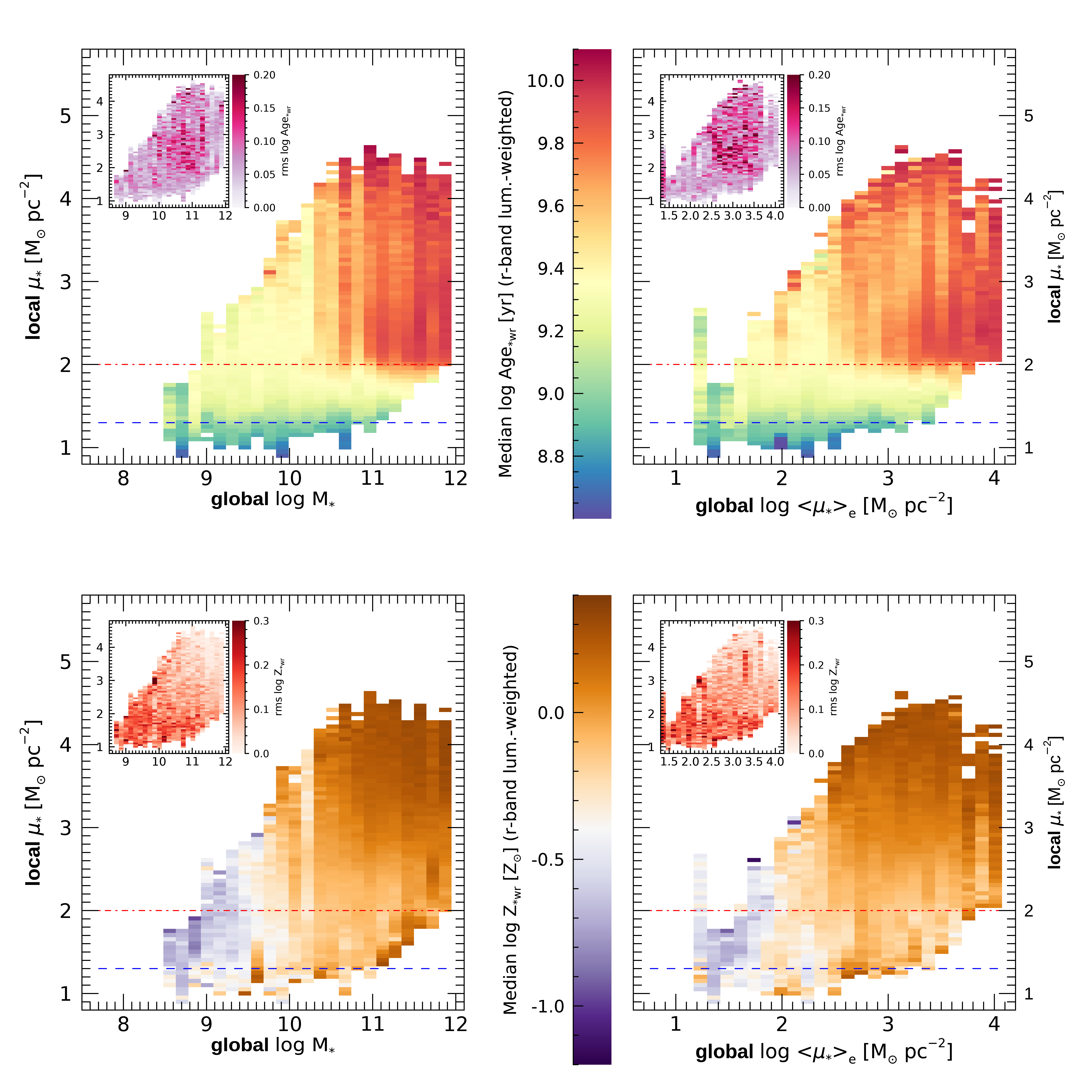}
    \caption{Bivariate dependence of local stellar age and metallicity on the local and global stellar mass (density). \emph{Top row}: median $\log$~\agewr~in bins of given $(M_*,\mu_*)$ (\emph{left panel}) and given $(\langle\mu_*\rangle_e,\mu_*)$ (\emph{right panel}). The insets show the r.m.s. deviation (half of the 16--84 percentile range) in each bin. \emph{Bottom row}: same as top row, but for $\log $~\zstarwr.
     The red dot-dashed line and the blue dashed line indicate the approximate completeness limit for the old-ridge and the young-sequence regions, respectively (see Sec. \ref{sub:SPpars}).}
    \label{fig:glocal_distr}
\end{figure*}

In Fig. \ref{fig:glocal_distr} the panels on the left display the distribution of regions in the $M_*$--\mustar~plane, colour-coded by the median \agewr~and median \zstarwr~in each 2D-bin, in the top and bottom row, respectively. The maps in the insets show the scatter in these two quantities in each $(M_*,\mu_*)$ bin. In the panels on the right, \mue~is considered instead of $M_*$ as a global parameter. These plots allow to appreciate how the \emph{median} \mustar--\agewr~relation and the \mustar--\zstarwr~relation change as a function of the global parameter, $M_*$ or \mue. Each column of the maps is, in fact, the median representation of such relations (with rms in the insets).

Starting with the top left panel of Fig. \ref{fig:glocal_distr}, we can quite clearly distinguish two regimes for galaxies more massive and less massive, respectively, than the threshold mass $M_{*,\rm{thr}}\sim 10^{10.3}\,\rm{M}_\odot$. Above $M_{*,\rm{thr}}$ the \mustar--\agewr~relation is approximately flat, with \agewr~values around $10^{9.8-9.9}$\,yr over the \mustar~range above $10^2$\,\msunpcsq. Further above $10^{11}\,\rm{M}_\odot$ the coverage of the \mustar~range below $10^2$ is highly incomplete because of the surface brightness cut and of the high $M/L$ implied by the old age and high metallicity. Nevertheless, in massive galaxies we observe a systematic trend towards ages younger than $10^{9.5}\,\rm{yr}$ at \mustar$\lesssim 10^2$. In the light of the age profiles of early-type galaxies (ETGs) presented in our previous work \citep{Zibetti:2020aa}, we can interpret the flat part of the \mustar--\agewr~relation as the contribution of ETGs and bulges, while the down-bending of the relation can be attributed to the disks of early-type spirals \citepalias[see also][]{Zibetti:2017aa}. Below $M_{*,\rm{thr}}$ we observe that \emph{i)} the highest \mustar~that can be reached diminishes with $M_*$ (see also Fig. \ref{fig:glocal_mass_corr}) and \emph{ii)} the lowest \mustar~regions ($\log $\mustar $< 1.5$) have systematically lower ages than the regions of higher \mustar, by approximately 0.25 dex. 
It is also interesting to point out that the scatter in \agewr~at fixed $(M_*,\mu_*)$ is larger in the intermediate mass range, across the transition mass.

\begin{figure*}
	\includegraphics[width=\textwidth]{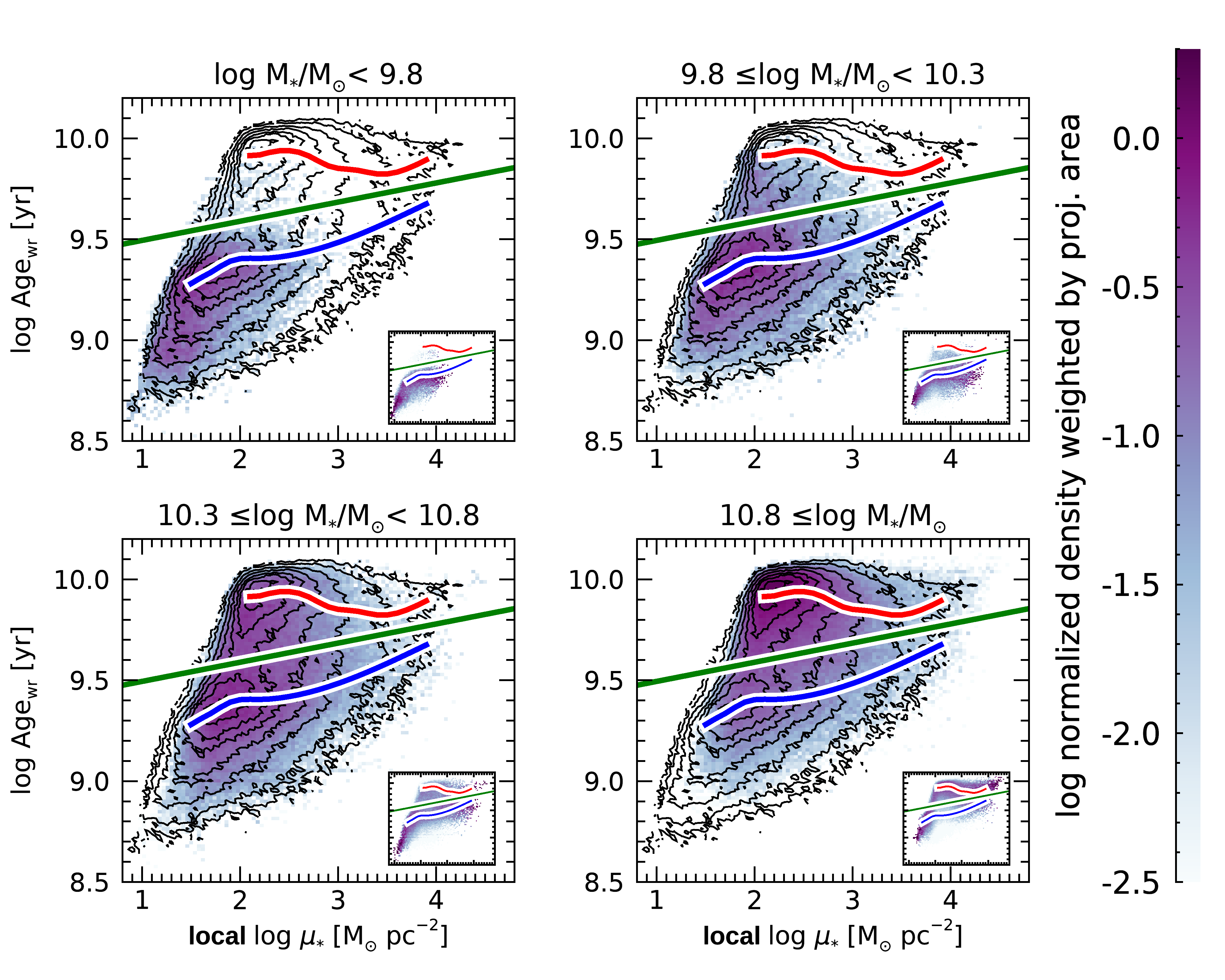}
    \caption{Distributions of regions weighted by physical projected area \agewr--\mustar~planes in four bin of stellar mass $M_*$. The density is normalized to the peak of the distribution for the full sample. Isodensity contours as well as the old ridge (red line), young sequence (blue line) and the green-valley divide (green line) are those for the full sample, the same as in the top left panel of Fig. \ref{fig:ageZ_mass_relation}. The insets of each panel represent the corresponding distributions after normalizing the integral of the distribution in each $x$-bin. A change in the balance between the young-sequence and the old-ridge populations is apparent as a function of $M_*$, yet the \mustar--\agewr~relations for the two populations do not appear to vary.}
    \label{fig:age_mustar_massbins}
\end{figure*}

It is worth reminding here that the distribution of \agewr~at fixed \mustar~is bimodal (see Fig. \ref{fig:ageZ_mass_relation}), so the shift in the median \mustar--\agewr~relation as a function of $M_*$ can be either due to an overall shift of the distribution(s) (viz. of the young sequence or of the old ridge) or to a different balance between the weight in the young sequence and in the old ridge, or to a combination of these two effects. Fig. \ref{fig:age_mustar_massbins} addresses this question by showing the \mustar--\agewr~distributions in four bins of $M_*$, with the same representation as in the top left panel of Fig. \ref{fig:ageZ_mass_relation}. The contours and the lines representing the young sequence, the old ridge and the green-valley division are those derived from the full sample distributions (all $M_*$). As a function of $M_*$ we can clearly see that the balance between the young-sequence population and the old-ridge population changes, with the former being dominant at small $M_*$ and the latter being dominant at large $M_*$. However, the positions of the peaks, when present, are consistent with the peaks from the full-sample distribution (see also Appendix \ref{app:oldyoung_massplanes}). This leads us to the conclusion that the systematic variation of the median \mustar--\agewr~relation as a function of $M_*$ is primarily driven by the change in the balance between the young-sequence and the old-ridge population, while keeping unchanged the \mustar--\agewr~relation for the two populations.

\begin{figure*}
\centering
	\includegraphics[width=0.48\textwidth]{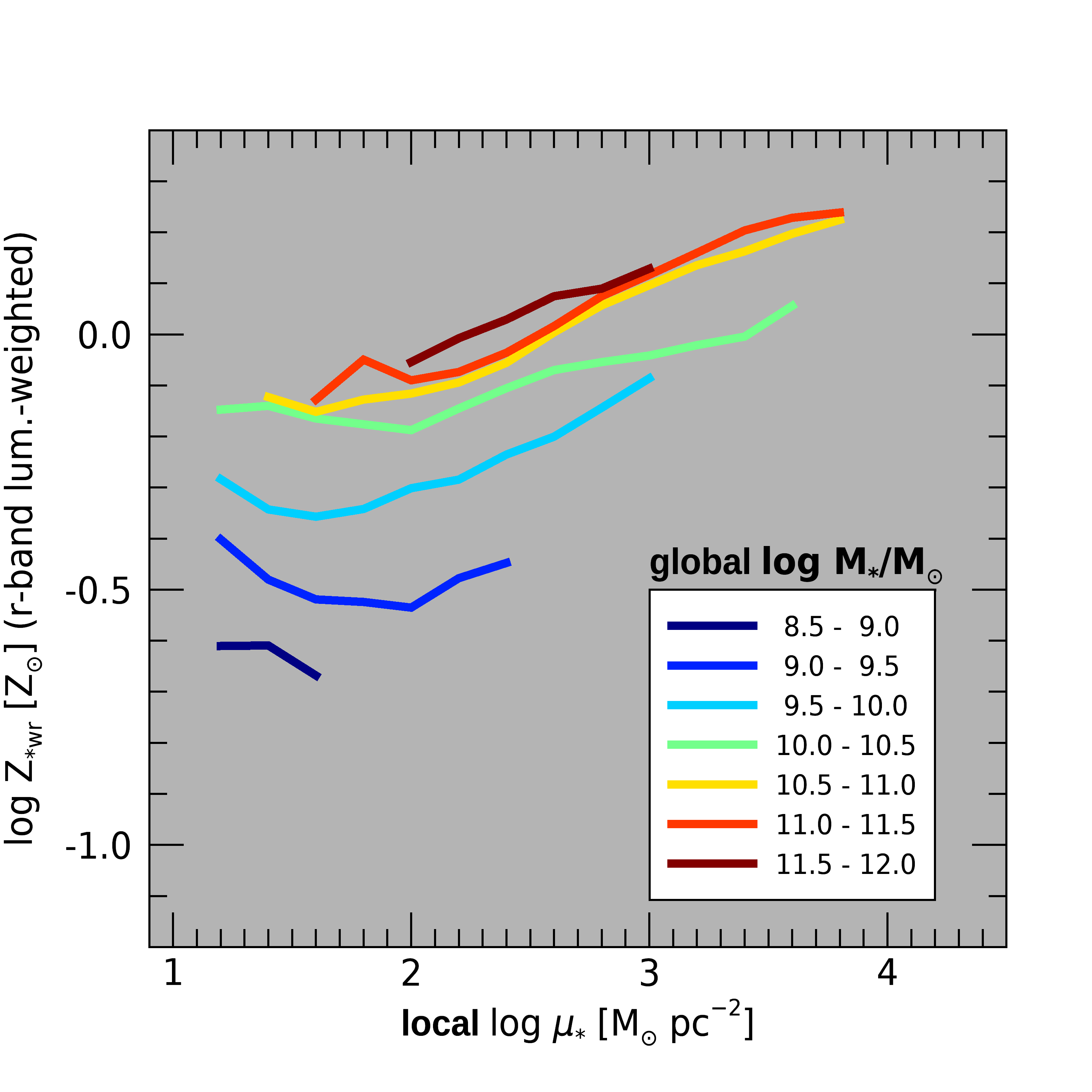}
	\includegraphics[width=0.48\textwidth]{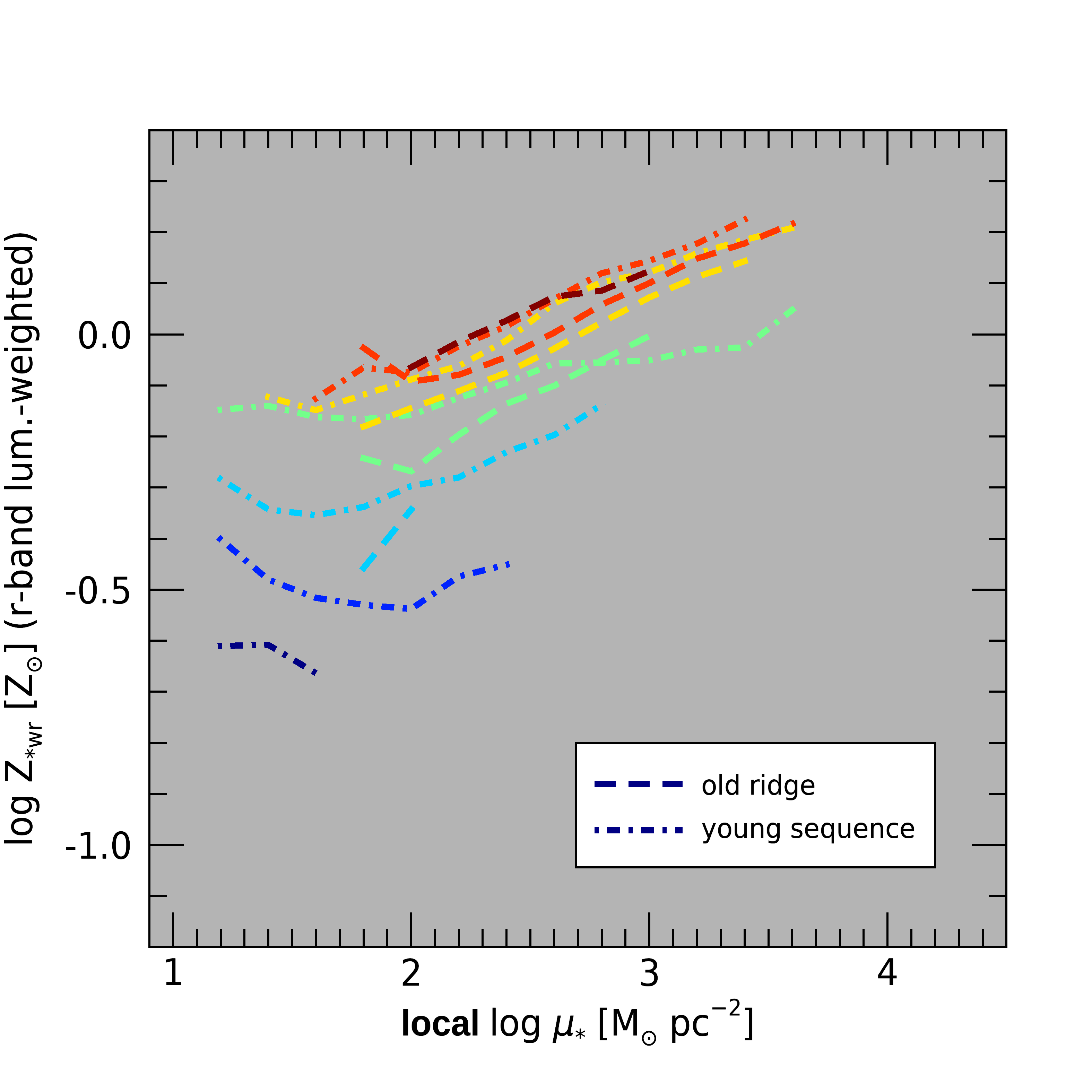}
    \caption{Median \mustar--\zstarwr~relations in bins of galaxy stellar mass $M_*$ (left-hand panel). In the right-hand panel, the $M_*$ bins have been further split into subsamples of regions belonging to the young sequence (dash-dotted lines) and to the old ridge (dashed lines), respectively.}
    \label{fig:z_mustar_massbins}
\end{figure*}

Looking at the metallicity \zstarwr~as a function of $(M_*,\mu_*)$ (bottom left panel of Fig. \ref{fig:glocal_distr}), the appearance of a dual distribution is also quite clear. Regions in galaxies more massive than $M_{*,\rm{thr}}$ are characterized by overall high metallicity \zstarwr$\gtrsim10^{-0.2}\rm{Z}_\odot$ and strong correlation between \zstarwr~and \mustar. These are, in fact, the typical properties of the ETGs that dominate this mass regime, as analyzed in detail in \citet{Zibetti:2020aa}. Less massive galaxies host regions that are \emph{on average} more metal-poor than those in massive galaxies. This is only partially a consequence of the lower \mustar~(hence lower \zstarwr) attained by low-mass galaxies. At fixed \mustar$\sim 10^2\,$\msunpcsq~we observe a clear trend for decreasing \zstarwr~with decreasing $M_*$. From the rms inset we can also note that in the high-mass/high-\mustar~regime the \mustar--\zstarwr~relation is extremely tight with a scatter of a few 0.01 dex, i.e. well below the level of statistical errors for our estimates, while the scatter goes up to $\sim 0.2$\,dex in the low-mass regime and in the low-\mustar~regime of massive galaxies (possibly the disks of early-type spirals). From this plot we can see that the global parameter $M_*$ does not only determine two different regimes, but also modulates the median \mustar--\zstarwr~relation within these two regimes. This can be seen even more directly in the left panel of Fig. \ref{fig:z_mustar_massbins}, where we plot the median \mustar--\zstarwr~relations in different bins of $M_*$. A systematic up-shift of the relation for increasing $M_*$ is apparent.

The right-hand panel of Fig. \ref{fig:z_mustar_massbins} displays the median \mustar--\zstarwr~relations in different bins of $M_*$, now split into young-sequence and old-ridge regions. In partial contrast to what we saw in Fig. \ref{fig:Z_mass_relation_bluered}, where the two subsamples shared an almost identical \mustar--\zstarwr~relation, we note that within a given mass bin and at fixed \mustar, young-sequence regions have systematically higher median \zstarwr~with respect to old-ridge regions by $\lesssim 0.1$ dex typically (provided that both subsamples are significantly populated in that $M_*$ bin). These small differences disappear almost completely when the full $M_*$ range is considered as a result of the multivariate correlations between \mustar, $M_*$, \agewr~and \zstarwr.

The two panels on the right of Fig. \ref{fig:glocal_distr} repeat the same analysis of the left-hand panels but using \mue~instead of $M_*$. The basic features of these plots resemble very closely those obtained with $M_*$. The main difference is the smoother transition from the low- to the high-\mue~regime, which occurs at $\langle\mu_*\rangle_e\sim 10^{2.5}\,$\msunpcsq. By looking at the insets, it appears that the scatter in \agewr~and \zstarwr~at fixed $($\mue$,\mu_*)$ is somewhat higher than at fixed $(M_*,\mu_*)$. This might indicate that, once \mustar~is fixed, the main ``second parameter'' that determines age and metallicity is $M_*$, while the effect of \mue~derives from its correlation with $M_*$. It is worth noting, though, that \mue~is known to less accuracy than $M_*$ by construction, as it incorporates uncertainties on both $M_*$ and $R_e$, and this artificially increases the observed scatter in age (metallicity) in a given (\mue, \mustar) bin above its intrinsic value. In fact, if we allow for a $\sim0.1$~dex error on $R_e$ (which is reasonable for nearby galaxies) this would turn into a $\sim0.2$~dex error to be summed in quadrature with the error on $M_*$. While we should consider that the gradient of variation of median age (metallicity) along \mue~is quite flat, an error in excess of 0.2 dex along this axis can certainly reshuffle the points and explain a significant portion (if not all) of the excess scatter. Our current level of uncertainties (and knowledge thereof) does not allow us to draw firm conclusions about the possible dominance of $M_*$ over \mue~as the ``second parameter'' in these relations.

\section{Discussion}\label{sec:discussion}
We can read the results presented in the previous Sections starting from a \emph{local} perspective. The local \mustar~broadly determines at first order the metallicity \zstarwr. The local \mustar~also determines the age \agewr~in case of regions sitting on the young sequence, while for old-ridge regions the age is broadly independent of \mustar. We found that the probability distribution for \zstarwr~at given \mustar~is largely independent of the region being on the old ridge or on the young sequence when the full range of host galaxy $M_*$ is considered (as we saw in Fig. \ref{fig:z_mustar_massbins}, however, the young-sequence regions tend to have slightly higher metallicity than old-ridge regions when $M_*$ is fixed in a limited range). The metal enrichment of stars is mainly determined by the local density that the region has been able to build up, with very little dependence on the time when this stellar mass has been formed. In other words, the local metallicity depends on the integral of the local SFH, but very little on its shape\footnote{The crucial role of local mass build-up on determining \zstarwr~was already pointed out by \cite{Zhuang:2019} based on dynamical \mustar~estimates, but in this work we explicitly quantify the dependence (or lack thereof) on the shape of the SFH, which was not analysed before.}. On the other hand, a region knows also about its global environment, i.e. its parent galaxy. The total stellar mass $M_*$, in first place, and the average stellar mass density \mue, possibly just by inherited correlation with $M_*$, determine a shift in the local scaling relations. By directly comparing local and global drivers in our work, we have been able to highlight the existence of a global $M_*$ threshold (corresponding also to a \mue\, threshold) that determines different regimes for the local relations. In the high-mass regime ($M_*\gtrsim 10^{10.3}\,\rm{M}_\odot\equiv M_{*,\rm{thr}}$) a region is much more likely to inhabit the old ridge, and it is bound to follow a very tight \mustar--\zstarwr~relation extending to super-solar values. Below the transition mass, most regions live in the young sequence and their distribution in \mustar~is shifted towards a factor $\sim 10$ lower density than in massive galaxies. Their \zstarwr~distribution is therefore shifted to lower metallicity as a consequence of both the general \mustar--\zstarwr~relation at these lower \mustar~and of the shift of the \mustar--\zstarwr~relation to lower values in the low-$M_*$ range.

Galaxy stellar mass $M_*$ is linked to (or even determines) \emph{i)} the highest \mustar\,that can be reached, \emph{ii)} the capability of fuelling SF for long time (i.e. the time-scale of the SFH), \emph{iii)} the ability of retaining metals globally. Locally \agewr~and \zstarwr~are tightly connected to \mustar.  On the one hand, this suggests that the capability to build-up \mustar~is intimately connected to the \emph{shape} of the SFH, hence to the resulting \agewr.  On the other hand, the tight \mustar--\zstarwr~relation, which is virtually independent of anything else above the threshold mass at $M_*\gtrsim 10^{10.3}\,\rm{M}_\odot$, indicates that in this high mass regime the stellar metallicity is mainly determined by the local gravitational potential, i.e. the ability of binding metals locally. Below the threshold mass, the global stellar mass (i.e. the depth of the global potential well) appears to become the main limiting factor to the growth of metallicity, as supported by the lack of a clear and universal \mustar--\zstarwr\, relation and, on the contrary, by an evident dependence of local \zstarwr\, on $M_*$ (see Fig. \ref{fig:z_mustar_massbins}). 

This interpretation is in substantial agreement with the scenario that \cite{Barone:2018,Barone:2020} propose in order to explain the relation between the \emph{global} stellar metallicity and gravitational potential in SAMI galaxies. In this view, the gravitational potential determines the ability of a galaxy to retain enriched gas and recycle it into new generations of stars. This is what \cite{Scott:2009} also pointed out based on their analysis of the \emph{local} escape velocity vs. metallicity relation in ETGs and is consistent with the basic expectations from models of self-regulated in-situ star formation \citep[e.g.][]{Pipino:2010}.

Very recent results by \cite{Neumann.2021}, which appeared while we were completing this manuscript, confirm the main trends concerning the relations between stellar metallicity, stellar mass and surface mass-density, based on a totally independent sample of 7439 MaNGA galaxies and on a different stellar-population analysis method (FIREFLY). In particular their work highlights the presence of a strongly significant local mass-density--metallicity relation, which becomes steeper and better defined at high global stellar masses, in agreement with our findings. Small yet significant systematic quantitative differences with respect to our results are apparent in the slope and the shape of the \mustar-\zstarwr~relations.

In this respect, it is worth spending some words of caution about the potential biases that may affect the complex analysis presented in this work. In fact, \cite{Gonzalez-Delgado:2014ab} conducted a study similar to ours on a sample of 300 CALIFA galaxies, aimed at assessing the role of $M_*$ and \mustar\, in determining the local metallicity of stars. Their analysis is based on the full-spectral fitting, full-inversion algorithm STARLIGHT \citep{STARLIGHT}. Their results partly agree with ours, in that both global stellar mass and local \mustar\, appear to have a significant role. However, they do not find evidence for two different regimes above and below $M_*\sim 10^{10.3}\,\rm{M}_\odot$ and show that also at high masses the global stellar mass determines local \mustar--\zstarwr\,relations that are differently offset relative to each other. The reasons for this discrepancy is currently unclear, but it is arguably related to the different approaches to stellar population properties estimates \citep[see discussion in][]{Zibetti:2020aa} and also to the different definition of mean metallicity (mean mass weighted logarithmic in their case vs. mean light-weighted linear in our case).

Turning to stellar age, the results by \cite{Gonzalez-Delgado:2014aa} also differ from ours, especially in the high mass regime, above $M_*\sim 10^{10.3}\,\rm{M}_\odot$. There we find no or very mild dependence of \agewr\, on either \mustar\, or $M_*$, while \cite{Gonzalez-Delgado:2014aa} found a very clear correlation between $\langle\log \mathrm{age}\rangle_L$ and both \mustar\,(mainly) and $M_*$ (see their section 8.4, figures 13 and 14). In the low mass regime, we note a better qualitative agreement with our results, in that the mean stellar age correlates mainly with \mustar\, and only marginally or with larger scatter with $M_*$.

\subsection{The mass-metallicity relation of passive and star-forming galaxies: from spatially resolved to global scaling relations}\label{subsec:peng2015}
As we discussed in Sec. \ref{sub:mass_age_all} and \ref{sub:mass_Z_all} by comparing the semi-resolved mass-age and mass-metallicity relations of Fig. \ref{fig:ageZ_mass_relation} with the analogous \emph{global} relations of \cite{gallazzi+05}, the translation of local relations into global ones is far from trivial and potentially affected by several biases, such as the aperture bias, the brightness bias (i.e. bright young regions outshining the old ones), the incomplete coverage of low \mustar\, regions. In particular it is relevant to stress that global measurements based on integrated spectra are effectively luminosity-weighted means, because an integrated spectrum is indeed the luminosity-weighted average of the spectra of all regions. As opposed, the spatially resolved relations are area-weighted and tend to be biased towards the properties of low surface mass-density regions, which represent the majority of the projected area of a galaxy. Investigating the possible local origin of global scaling relations requires careful consideration of these observational effects, which are driven by real structural differences in galaxies. On the other hand, we can take advantage of this strong link with structure to shed light on the physics that governs scaling relations from a different perspective than the commonly adopted perspective of a global baryonic cycle.

We focus on the mass-metallicity relation (MZR hereafter) and on the differences between the relation followed by star-forming and passive galaxies, respectively. \cite{peng_maiolino_cochrane2015} used the stellar metallicity of 26\,000 SDSS galaxies from \cite{gallazzi+05} and discovered the presence of two distinct MZRs: a flatter relation with higher normalization for passive galaxies and a steeper relation for the star-forming galaxies, which converges to the passive relation at $\log (Z_*/Z_\odot) \sim 0.15$ for $M_*\sim 10^{11}\rm{M}_\odot$. This result was confirmed by \citet{Trussler:2020aa,Trussler:2021aa}, using a different stellar population analysis, and by \cite{Gallazzi:2021aa}. Both groups highlighted that the difference between passive and star-forming galaxies is the fundamental driver of the scatter in the MZR, largely independent of the external conditions (e.g. environment in which galaxies live). 

\cite{peng_maiolino_cochrane2015} interpret the higher metallicity of passive galaxies as a consequence of ``strangulation'' mechanisms that make galaxy evolve from star-forming to passive. While normal star-forming galaxies are continuously replenished of fresh metal-poor gas that sustains the SFR, passive galaxies are created by removing this external reservoir via ``strangulation'' (or better say ``starvation''). In this way, the last generations of stars are formed entirely from the highly enriched gas that is available inside the galaxy, thus significantly increasing the average stellar metallicity, like in a classic closed-box model. \citet{Trussler:2020aa} performed more in-depth analytical modeling based on the ``gas regulator'' model by \cite{Peng-Maiolino:2014ab} and came to the conclusion that both starvation and outflows drive the transformation of galaxies from star-forming to passive and the observed bimodality in the MZR.

\cite{Spitoni:2017aa} have also modeled the two MZRs with analytical models of metal enrichment, that assume exponential gas infall and star-formation-driven, metal-loaded outflows. Interestingly, they show that the lower metallicity of star-forming galaxies with respect to equal-mass passive galaxies results from their extreme inefficiency in converting the infallen gas into stars and, at the same time, from the strength of their winds. In fact, according to this model, star-forming galaxies along their history must have received a factor 10 to 100 more mass from gas infall than they currently have in stars. This factor is just a few for currently passive galaxies, as opposed. The difference is tightly related to the different time-scale for the infall (short for passive galaxies and long for star-forming ones) and to the different strength of the ``wind factor'', which couples the SFR to the outflow mass (weak for passive galaxies, strong for star-forming ones). \\

These works explain the diversity of MZRs in terms of the \emph{global} baryonic cycle of galaxies. In the present paper, on the other hand, we have shown that the spatially-resolved \mustar-$Z_*$ relation is totally consistent with being the same for young and old regions when galaxies of all $M_*$  are considered, while small differences emerges at fixed $M_*$, in the sense of young regions being slightly more metal-rich than the old ones at fixed \mustar. This is in seeming contradiction with the findings for the global MZR. We note that in \citeauthor{gallazzi+05} (\citeyear{gallazzi+05}, hence in \citealt{peng_maiolino_cochrane2015}) and in this work, \zstarwr~is derived employing similar methods and observational constraints and using very similar SSP models as the base of the model libraries, and moreover quantity definitions are fully consistent. Thus we can expect systematic effects \emph{not} to be the main source of this apparent contradiction. Its origin is rather to be identified in the different structure (i.e. stellar mass distribution) of star-forming and passive galaxies. In fact, \emph{at a given $M_*$}, passive galaxies are on average more compact and denser than star-forming ones \citep[e.g.][]{blanton_etal03,Williams:2009aa}. This means that the average \mustar\,of passive galaxies is higher than that of star-forming galaxies, and hence sample the \mustar--\zstarwr~relation over a different range of \mustar. Therefore, the mean metallicity of a passive galaxy must be \emph{on average} higher than the metallicity of a star-forming galaxy of equal mass. Interestingly, the diverging trends of the global mass-metallicity relation for the two classes of galaxies mirror the diverging trends of the mass-size relation \citep{shen_etal03}, hence of the mass--mean-surface-density relation. 

Beside this prominent structural effect, we must consider a luminosity bias that affects the global metallicity estimates in star-forming galaxies. Low-\mustar\, regions are characterized by both lower age and lower metallicity, which result in lower mass-to-light ratio: these regions, therefore, tend to be relatively over-luminous and bias the average metallicity (derived from the integrated spectrum) to lower values with respect to a properly mass-weighted mean of the different regions. In addition, the ``metal dilution'', which results from the infall of metal-poor gas into star-forming regions, may also boost the differences between the two MZRs to some extent. However, as we noted before, in a given bin of $M_*$, the young regions have systematically \emph{higher} metallicity than older regions at fixed \mustar, a small effect ($\lesssim 0.1$~dex), yet actually opposite to what ``metal dilution'' would imply. This suggests that, if anything, such a mechanism is overall not so strong to significantly bias the metallicity estimates on global scale. Although none of the aforementioned biases can be accurately quantified at the moment, \emph{it appears that structural differences are the most likely driver for differences in global metallicity as large as $0.3-0.4$\,dex between passive and star-forming galaxies, as shown by \cite{peng_maiolino_cochrane2015}, in presence of the observed \mustar--\zstarwr\, relation}.

Finally, we note that ``aperture effects'', caused by the finite size of the SDSS fibres through which the integrated galaxy spectra are collected, cannot play a major role in determining the duality of the MZR. Since radial metallicity gradients are in general negative (as testified by the \mustar--\zstarwr\, relation) and star-forming galaxies are larger (on average) than passive galaxies of the same mass, the metallicity estimates for star-forming galaxies should be more biased than for passive galaxies towards the larger metallicity values that characterize the inner regions. So, if anything, correcting for aperture biases may increase the observed differences between the two MZRs, rather than decrease them.

In summary, our results and considerations in this Section indicate that the global baryonic cycle and the overall metal enrichment of a galaxy is tightly connected to its structure and to the ability to accumulate and retain metals locally.  
The existence of a well defined \mustar--\zstarwr\, relation on local scales, although partly modulated by $M_*$ and local age, suggests that there is an underlying physics, common to all galaxies, that describes how stellar density and metallicity build up locally. Yet it is the global physics (e.g. overall availability and efficiency of gas accretion, overall outflow strength etc.) that determines the structure and the timescale of the star formation and assembly.

\section{Summary and conclusions}\label{sec:summary}
In this work we have analyzed the dependence of the \emph{local} mean (light-weighted) stellar age (\agewr) and metal enrichment (\zstarwr) on both \emph{local} stellar-mass surface density (\mustar), and \emph{global} stellar mass ($M_*$) and mean stellar-mass surface density (\mue). We have studied bivariate (\agewr~and \zstarwr~vs. a single mass parameter) and three-variate distributions (\agewr~and \zstarwr~vs. two different mass parameters) for a sample of more than 600 thousand individual regions from 362 CALIFA galaxies. Our main findings can be summarized as follows:
\begin{itemize}
    \item The local age and metallicity of individual regions follow well defined relations with the local stellar-mass surface density. Specifically, the age \agewr~follows either a monotonically increasing relation with \mustar~(``young sequence'') or a flat relation around old ages (``old ridge''); the metallicity \zstarwr~follows a single, yet significantly dispersed, relation, monotonically increasing with \mustar.
    \item The \mustar--\zstarwr~relation is overall independent of regions belonging to the young sequence or to the old ridge, however a mild dependence appears at fixed $M_*$, with young regions being slightly more metal enriched (by $\lesssim 0.1$~dex) than older ones at given \mustar.
    \item Global mass parameters, $M_*$ and \mue, determine \emph{i)} a shift in the distribution of regions in \mustar~(the maximum \mustar~reached in a galaxy scales with $M_*$) and \emph{ii)} a shift in the local relations below a critical mass $M_*\sim 10^{10.3}\mathrm{M}_\odot$ (local relations are offset towards lower \agewr~and \zstarwr~with respect to massive galaxies). We found a tentative indication that $M_*$ may be predominant over \mue.
    \item Global mass--age and mass--metallicity relations \citep[e.g.][]{gallazzi+05} are only broadly reproduced by the distribution of regions in the planes of total $M_*$ vs. local \agewr~and local \zstarwr, respectively. We identified the possible origins of these differences in the structure of galaxies and in observational effects. 
    \item The stark contrast between the different global mass-metallicity relations of quiescent and star-forming galaxies \cite{peng_maiolino_cochrane2015}, on one side, and the lack of bimodality of the local \mustar--\zstarwr~relation, which is common to both young-sequence and old-ridge regions, on the other side, highlight the structural origin of such a dichotomy on global scales.
\end{itemize}
In our discussion we have highlighted how \emph{local stellar population properties are driven by both the local stellar-mass surface density} (largely irrespective of the time-scale on which this has been built up, as far as metallicity is concerned) \emph{and the global stellar mass}. In this sense we can conclude that stellar mass is a ``glocal'', i.e. simultaneously local and global, driver of the stellar population properties. The effect of the global stellar mass seems to saturate at $M_*\gtrsim 10^{10.3}\mathrm{M}_\odot$, above which the local \mustar~appears as the primary (or even only) driver of the stellar population properties. At lower masses, $M_*$ determines the structure of a galaxy and, in turn, the distribution of the individual regions on the local relations.

This leaves us with the next fundamental open question and challenge, that is how to integrate the interpretation of the global relations in terms of baryonic cycle with the structural relations that we have analyzed in this work.

\section*{Data availability}
The IFS datacubes on which this work is based are publicly available from the CALIFA webserver \texttt{https://califaserv.caha.es}. The maps of stellar population properties will be shared on reasonable request to the corresponding author.
\section*{Acknowledgements}
We thank the anonymous referee for the constructive and insightful report that helped improve the clarity of the paper and assess the robustness of our results.\\
S. Z. and A. R. G. have been supported by the EU Marie Curie Career Integration Grant ``SteMaGE'' Nr. PCIG12-GA-2012-326466  (Call Identifier: FP7-PEOPLE-2012 CIG).
%%%%%%%%%%%%%%%%%%%%%%%%%%%%%%%%%%%%%%%%%%%%%%%%%%



\bibliographystyle{mnras}
\bibliography{all_mybibs} % if your bibtex file is called example.bib

%\begin{thebibliography}{99}
%\end{thebibliography}

%%%%%%%%%%%%%%%%%%%%%%%%%%%%%%%%%%%%%%%%%%%%%%%%%%

%%%%%%%%%%%%%%%%% APPENDICES %%%%%%%%%%%%%%%%%%%%%

\appendix
\section{Effects of adaptive smoothing on the IFU cubes}\label{app:azmooth3}
As reported in the main text and in our previous works on CALIFA data \citep{Zibetti:2017aa,Zibetti:2020aa}, in order to enhance the signal-to-noise ratio ($S/N$) of the low-surface-brightness regions up to a level that allows a reliable measurement of the spectral indices, we adopt the adaptive smoothing algorithm originally developed in \textsc{adaptsmooth} by \citet{adaptsmooth} and \cite{ZCR09}. The algorithm replaces the spaxels where the $S/N$ does not reach a given target $S/N_\text{target}$, with the average (median) of a number of surrounding spaxels within a kernel, up to a maximum number (or distance) fixed by the user. The size of the kernel is adaptive, in that the minimum size that produces the target $S/N_\text{target}$ is used. On the one hand, this allows us to retain the maximum spatial resolution where no $S/N$ enhancement is required, but, on the other hand, in regions where smoothing is required, it introduces \emph{i)} variable spatial resolution and \emph{ii)} statistical correlation among the spaxels. Here we discuss whether this can lead to biases in our results and if the more popular approach of Voronoi binning the data \citep[e.g.][]{cappellari_copin_2003} could be better for the goals of our analysis.

First of all, we consider the fraction of spaxels that are affected by smoothing to some degree. Over a total of $601\,282$ spaxels, $51\,721~(8.6\%)$ have a level-2 smoothing applied (i.e. the median is taken among the spaxel itself and a corona of 1-spaxel width); $61\,497~(10.2\%)$ have a level-3 smoothing (the spaxel itself and a corona of 2-spaxel width); $46\,708~(7.8\%)$ are level-4 and $222\,646~(37.0\%)$ are level-5. More than one third of the spaxels, $218\,710~(36.4\%)$ are unaffected by the smoothing algorithm. More than half of the spaxels are smoothed with a kernel of no more than 2-spaxel radius.

\begin{figure*}
\centering
	\includegraphics[width=\textwidth]{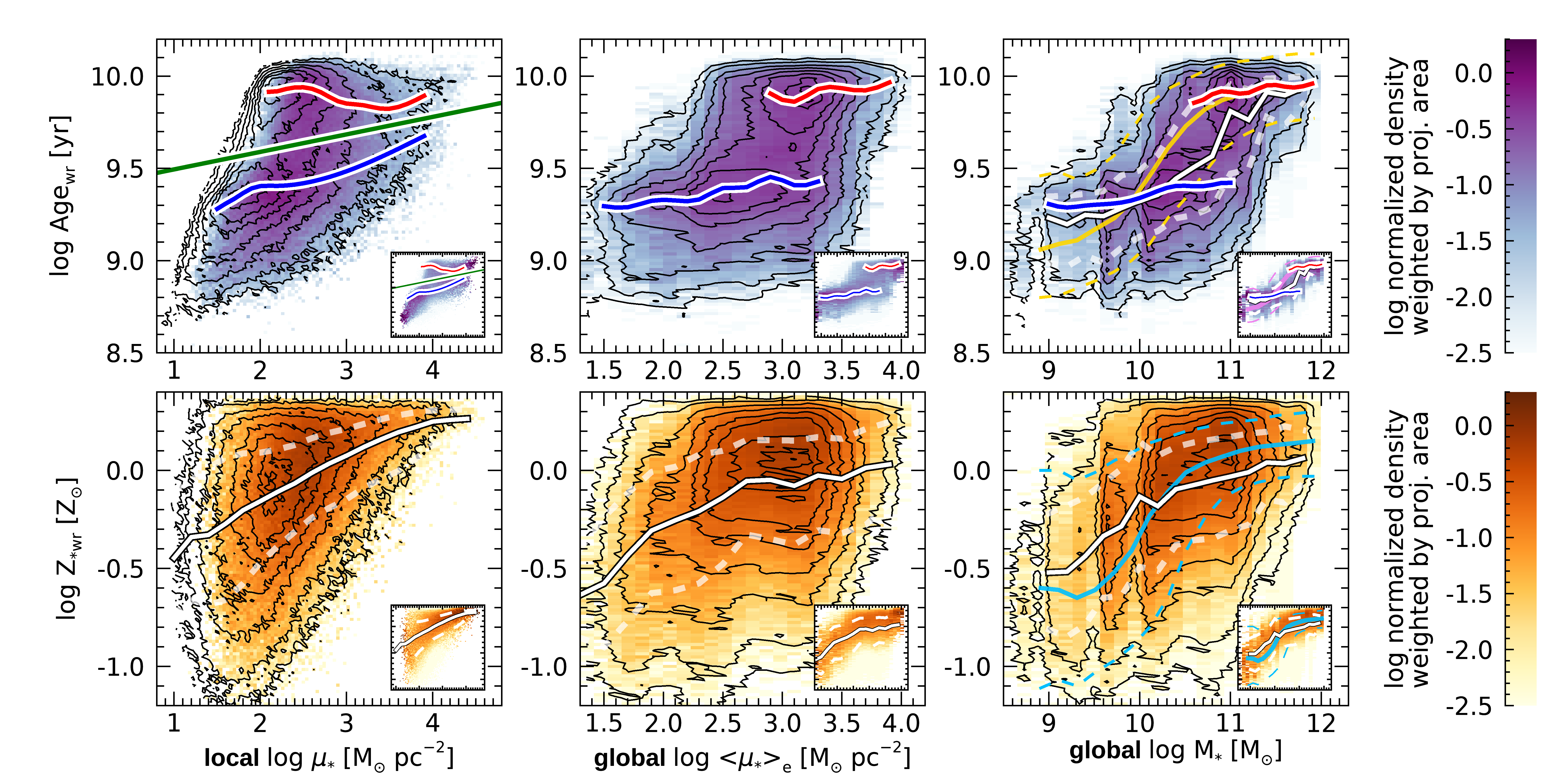}
    \caption{Distributions of regions weighted by physical projected area and normalized to the peak \emph{considering only the spaxels that are not affected by smoothing}, in the \agewr--mass (density) planes (\emph{top row}) and in the \zstarwr--mass (density) planes (\emph{bottom} row). In each row, from left to the right, the $x$ axis represents the \emph{local} stellar mass density \mustar, the \emph{global} mean stellar mass density within 1 effective radius \mue, and the \emph{global} stellar mass $M_*$. This figure is analogous to Fig. \ref{fig:ageZ_mass_relation} and adopt the same colour scales. The lines and contours are reported for reference from Fig. \ref{fig:ageZ_mass_relation} (i.e. they are relative to the full sample).}
    \label{fig:ageZ_mass_rel_unsmoothed}
\end{figure*}

In order to check for the impact of smoothed spaxels on our results, we have produced Fig. \ref{fig:ageZ_mass_rel_unsmoothed}, analogous to Fig. \ref{fig:ageZ_mass_relation}, which displays the distribution of regions in the different mass (density) vs. age planes and mass (density) vs. metallicity planes, using only the $36.4\%$ of spaxels that are not affected by smoothing. In this figure, lines and contours are the same as in Fig. \ref{fig:ageZ_mass_relation} (i.e. they refer to the full sample) and are reported as a reference. The colour scale is also the same as Fig. \ref{fig:ageZ_mass_relation}, i.e. same colour represents the same density of regions, properly weighted by projected area and sample completeness. Looking at the parameter spaces involving local \mustar~(left panels), we see that the effect of discarding the smoothed spaxels is essentially to move the low-\mustar~cut to the right by almost 1 dex, as expected. Apart from that, there is no noticeable change in the positions of the peaks for the \mustar--\agewr~relations nor in the median and percentiles of the \mustar--\agewr~relations, as it can be seen best in the insets with normalization in the \mustar~bins. Differences with respect to the full sample are apparent only when comparing the distributions as a function of the global parameters \mue~and $M_*$. In particular, the metallicity distributions at high $M_*$ and high \mue~are biased to higher values in the ``unsmoothed`` sample relative to the full sample. This effect, however, is not indicative of a bias of the method, rather is a mere consequence of cutting the sample and exclude the low-\mustar regions. In fact, the strong metallicity gradients observed in the massive galaxies translate a cut in \mustar into a cut in \zstarwr, thus resulting in \zstarwr distributions biased to higher values. On the other hand, despite these unavoidable offsets due to the different region selection, the main trends that are observed for the full sample are qualitatively reproduced by the unsmoothed sample too.

This a-posteriori check demonstrates that our results do not depend on the ``signal enhancement'' strategy that we have chosen for the low-\mustar~regions. In addition, we would like to stress that there are also good reasons of principle for choosing an adaptive smoothing rather than the commonly adopted Voronoi binning, especially for the analysis we have been running in our series of works on CALIFA. In fact, with the Voronoi tessellation one obtains \emph{statistical independence among the tiles}, but \emph{for the regions} (thereof there can be many of size $\sim 0.1$~kpc in a single Voronoi tile) \emph{the correlation is maximal}, actually all regions (spaxels) inside a Voronoi bin are treated as a single uniform region, thus completely erasing  any distribution and reducing it to a delta Dirac. When we produce the distributions presented in this paper, if we adopted the Voronoi tessellation we would just take one single value of \zstarwr, \agewr~and \mustar~and assign it a weight equal to the area of the tile. The adaptive smoothing allows to trace the spatial variations and give a much better representation of the width of the distributions of the parameters over the same area. While it is obvious that the (adaptive) smoothing produces some correlation among the spaxels, the degree of correlation that the Voronoi binning introduces is in any case larger, actually it is the maximum possible correlation.
As a matter of fact, \cite{adaptsmooth} shows that our adaptive smoothing algorithm preserves the total flux to very high accuracy and the spatial structure as well \citep[see figures in][]{adaptsmooth}. Moreover, in most cases the regions with stark spatial variations (e.g. strong radial gradients in surface brightness or stellar population properties) occur at relatively high surface brightness, in a regime that is left untouched by the \emph{adaptive} smoothing.

\section{Local and global mass (density) drivers for young-sequence and old-ridge regions}\label{app:oldyoung_massplanes}

In this appendix we illustrate in further detail the bivariate dependence of local stellar age and metallicity on the local and global stellar mass for old-ridge and young-sequence regions, separately. In particular, we repeat the analysis of Fig. \ref{fig:glocal_distr} and display the median $\log$~\agewr~and $\log$~\zstarwr~in bins of $(M_*,\mu_*)$, for the two sub-samples (left and right panels, respectively). By comparison with Fig. \ref{fig:glocal_distr}, we note that the median relations for the full sample are dominated by old-ridge regions above $M_{*,\text{thr}}\sim 10^{10.3}\text{M}_\odot$, and by young-sequence regions in the lower mass regime. In fact, the relations for the two subsamples are almost fully mirrored in the relations for the full sample in those two $M_*$ regimes, respectively. 

\begin{figure*}
	\includegraphics[width=\textwidth]{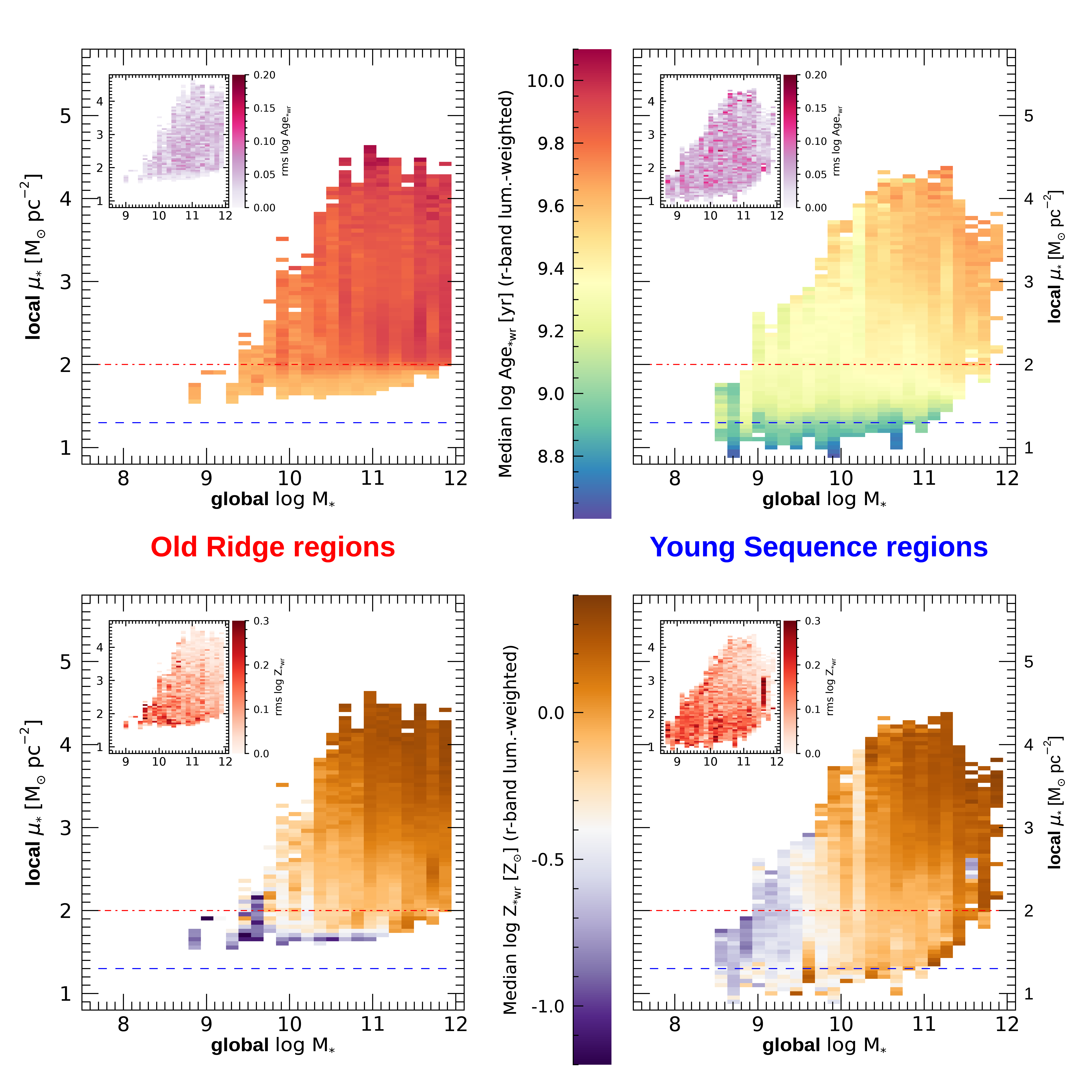}
    \caption{Bivariate dependence of local stellar age and metallicity on the local and global stellar mass.
    This figure is analogous to the \emph{left} column of Fig. \ref{fig:glocal_distr}, but for old-ridge regions (\emph{left-hand side plots}) and young-sequence regions (\emph{right-hand side plots}) separately. \emph{Top row}: median $\log$~\agewr~in bins of given $(M_*,\mu_*)$. The insets show the r.m.s. deviation (half of the 16--84 percentile range) in each bin. \emph{Bottom row}: same as top row, but for $\log $~\zstarwr. The red dot-dashed line and the blue dashed line indicate the approximate completeness limit for the old-ridge and the young-sequence regions, respectively (see Sec. \ref{sub:SPpars}).}
    \label{fig:glocal_distr_oldyoung}
\end{figure*}

In terms of \agewr, as expected, selecting a single age subsample results in a much smaller scatter with respect to the full sample. In particular, for the old-ridge regions \agewr~appears almost lack any dependence on $(M_*,\mu_*)$, for $M_*>M_{*,\text{thr}}$ and \mustar~above the completeness limit of $10^2$\msunpcsq, with a scatter within a bin of $\sim 0.05$~dex. Below the transition mass $M_{*,\text{thr}}$, the age trends of the young sequence drive the overall trends. The scatter for the young-sequence regions only is typically $\lesssim 0.1$~dex, throughout the plane and across the transition mass. This is in stark contrast with the scatter for the full sample, which increases up to $0.15-0.20$~dex in correspondence to the transition mass. This confirms that the overall trends and scatter are the result of the superposition of the two well-defined relations for the old ridge and the young sequence.

In terms of \zstarwr, the two subsamples display very similar relations, which confirms the analysis of Fig. \ref{fig:Z_mass_relation_bluered} and \ref{fig:z_mustar_massbins}. The scatter is everywhere dominated by the young-sequence regions.

% Don't change these lines
\bsp	% typesetting comment
\label{lastpage}
\end{document}